\documentclass[amssymb,prb,twocolumn,superscriptaddress,floats,showkeys,showpacs]{revtex4-2}
\usepackage[T1]{fontenc}
\usepackage{authoraftertitle}
\usepackage{bm}
\usepackage{graphicx}
\usepackage{amssymb}
\usepackage{leftidx}
\usepackage{upgreek}
\usepackage{siunitx}
\usepackage{amsfonts}
\usepackage{amsmath} 
\usepackage{hyperref} 
\usepackage{multirow}
\usepackage{makecell}
\usepackage{xcolor}
\usepackage{xspace}
\usepackage{comment}
\usepackage{xr}
\usepackage{tabularx}
\usepackage{adjustbox}
\usepackage{newtxmath}
\hypersetup{colorlinks,citecolor=blue, filecolor=blue ,linkcolor=blue , urlcolor=blue, pdftex}

\begin{document}

\newcommand{\HfS}{$\text{HfS}_{2}$\xspace}

\title{Pressure-induced optical anisotropy of \HfS}


\def \FUW{ Faculty of Physics, University
of Warsaw, Pasteura 5, 02-093 Warsaw, Poland}
\def \LNCMI{Laboratoire National des Champs Magnétiques Intenses, CNRS-UGA-UPS-INSA-EMFL, Grenoble, France}
\def \China{Hefei Innovation Research Institute, School of Microelectronics, Beihang University, Hefei 230013, P. R. China}
\def \CENT{Centre of New Technologies, University of Warsaw, Banacha 2c, 02-097 Warsaw, Poland}
\def \Wroclaw{Department of Semiconductor Materials Engineering, Faculty of Fundamental Problems of Technology, Wrocław University of Science and Technology, Wybrzeże Wyspiańskiego 27, 50-370, Wrocław, Poland}
\def \Spain{Geosciences Barcelona (GEO3BCN), CSIC, Lluís Solé i Sabarís s.n., 08028, Barcelona, Catalonia, Spain}
\def \MagdaG{Department of Materials Science and Engineering, National University of Singapore, 117575, Singapore} 
\def \MagdaGk{Institute for Functional Intelligent Materials, National University of Singapore, 117544, Singapore}

\author{Igor Antoniazzi} 
\email{igor.antoniazzi@fuw.edu.pl}
\affiliation{\FUW}
\author{Tomasz Woźniak}
\affiliation{\FUW}
\author{Amit Pawbake}
\affiliation{\LNCMI}
\author{Natalia Zawadzka}
\affiliation{\FUW}
\author{Magdalena Grzeszczyk}
\affiliation{\FUW}
\affiliation{\MagdaGk}
\author{Zahir Muhammad}
\affiliation{\China}
\author{Weisheng Zhao}
\affiliation{\China}
\author{Jordi~Ib{\'a}{\~n}ez}
\affiliation{\Spain}
\author{Clement Faugeras}
\affiliation{\LNCMI}
\author{Maciej R. Molas}
\affiliation{\FUW}
\author{Adam Babiński}\email{adam.babinski@fuw.edu.pl}
\affiliation{\FUW}

\begin{abstract} 
The effect of pressure on Raman scattering (RS) in the bulk \HfS is investigated under hydrostatic and non-hydrostatic conditions.
The RS lineshape does not change significantly in the hydrostatic regime, showing a systematic blueshift of the spectral features.
In a non-hydrostatic environment, seven peaks emerge in the spectrum ($P$=7~GPa) dominating the lineshape up to $P$=10.5~GPa.
The change in the RS lineshape manifests a pressure-induced phase transition in \HfS.
The simultaneous observation of both low-pressure (LP) and high-pressure (HP) related RS peaks suggests the corresponding coexistence of two different phases over a large pressure range.
We found that the HP-related phase is metastable, persisting during the decompression cycle down to $P$=1.2~GPa with the LP-related features finally recovering at even lower pressures.
The angle-resolved polarized RS (ARPRS) performed under $P$=7.4~GPa revealed a strong in-plane anisotropy of both the LP-related A$_{1g}$ mode and the HP peaks.
The anisotropy is related to the possible distortion of the structure induced by the non-hydrostatic component of the pressure.
We describe the obtained results by the influence of the non-hydrostatic pressure on the observed phase transition.
We interpret our results in terms of a distorted $Pnma$ phase as a possible HP induced structure of \HfS.
\end{abstract}


\date{\today}

\maketitle



\section{Introduction \label{sec:Intro}}

Transition metal dichalcogenides (TMDs) have arisen as an exciting class of materials pursuing a layered van der Waals (vdW) structure.
Their unique properties, which are strongly dependent on the structure thickness and the relatively easy exfoliation, have become a hot topic of material research.
Moreover, their layered structure makes them very sensitive to the interlayer spacing, which can be modulated by temperature or strain.
This explains the interest in the strain engineering of TMDs as an effective way to modify their characteristics.~\cite{DENG201814, Pimenta2023, PEI2022, Zhang2020}
In this study, we address the properties of hafnium disulfide (\HfS), a less investigated member of the TMD family.
The interest in the material is triggered by its promising electrical properties.
According to theoretical calculations, the room-temperature electron mobility in \HfS is much higher than in the most explored MoS$_2$.~\cite{Zhang2014}.
Few-layer \HfS field-effect transistors fabricated by Kanazawa et al.~\cite{kanazawa2016} are characterized by high drain currents and mobility with a transistor on/off current ratio larger than 10$^4$, good responsivity ($\sim$1.6 $\mu$AW$^{-1}$) in a transfer-free photodetector,~\cite{jitendra2022} and is quoted to the development of thermoelectric and optoelectronic devices.~\cite{deobrat2019}
The properties of \HfS are also highly susceptible to changes in temperature and pressure.~\cite{peng2021, ibanez2018high, Grzeszczyk2022, Zawadzka2023, antoniazzi2023}.
However, there is an ongoing debate regarding the precise pressure at which the pressure-induced phase transitions occur, whether the transitions are reversible upon pressure release, and what the crystallographic structure of the pressure-induced phase is.
Under ambient conditions, \HfS belongs to the space group $P\Bar{3}m1$ (No. 164).
The onset of the phase transition was reported at $P\sim$11 GPa based on Raman scattering (RS) measurements in Ref. ~\cite{ibanez2018high}.
The RS spectroscopy of \HfS allowed us earlier to report on two phase transitions~\cite{Grzeszczyk2022}, which were observed at $P\sim$9.8~GPa and $P\sim$15.2~GPa. 
The specificity of our previous study was the emergence of several new high-pressure (HP) peaks in the RS spectra.
A reversible transformation to the 3D orthorhombic $Immm$ (No. 71) structure at approx 12~GPa accompanied by semiconductor-to-semiconductor transition was reported in Ref.~\cite{zhong2023} mediated by neon gas as the pressure transmitting medium (PTM). 
Notably, the authors observed that the RS measurements reflected the transition under pressure as low as $\sim$ 9.2~GPa, which corresponds to other reports.~\cite{hong2022, Zhang2023}
Hong et al.~\cite{hong2022} reported two structural phase transitions at 8.0~GPa and 15.2~GPa as well as metallization at 20.5~GPa unsing helium as the PTM.
They also showed that upon decompression the characteristic RS peaks, which appeared above the transformation, were preserved down to 1.4~GPa pointing out large pressure hysteresis effect for \HfS.
Quite recently, Zhang et al.~\cite{Zhang2023} also reported two structural transitions around $P$=11.0~GPa and 35.5~GPa using silicone oil as  the PTM. 
The former transformation has been related to a transition to the orthorhombic $Pnma$ (No. 62) phase, and the latter to a structural transition from the $Pnma$ phase to the tetragonal $I4/mmm$ (No. 139) phase.
The pressure-induced transition from the $P\Bar{3}m1$ to $Pnma$ phase under pressure was also supported by first-principles calculations.~\cite{Zhang_2022}

To add more information on the structural changes of \HfS upon compression, we report on RS measurements as a function of pressure up to $P$=10.5~GPa.
We focus our work on the effect of the pressure hydrostaticity on phase transition by examining the results of experiments performed with two PTM.
No apparent phase transition is observed while using the 4:1 methanol-ethanol (ME) mixture, providing hydrostatic conditions of pressure within the investigated pressure range.
On the contrary, a substantial change in the RS spectrum of \HfS is observed under approx. $P$=7~GPa with a crystallographic oil as PTM.
Similarly to the observation by Hong et al.~\cite{hong2022}, a large hysteresis effect is observed and the phase transition is shown to be reversible upon full decompression.
The angle-resolved polarized RS (ARPRS) measurements performed at $P$=7.4~GPa revealed a strong in-plane anisotropy of the observed RS modes.
We discuss the possible origin of the anisotropy.
Our results point out the crucial effect of non-hydrostaticity on the phase transitions in \HfS under external pressure.

\section{Experimental results\label{sec:Results}}

\subsection{Raman scattering at ambient conditions} \label{ss:Pdep_RTemp}

There are 6 vibrational modes present in the total representation~\cite{lukovsky1973IR} under ambient conditions in \HfS with the 1T structure:

\vspace{-20 pt}
\begin{center}
\begin{equation}
\Gamma=\textrm{A}_\textrm{1g} + \textrm{E}_\textrm{g} + 2\textrm{A}_\textrm{2u} + 2\textrm{E}_\textrm{u}
\end{equation}
\end{center}

\begin{figure}[h]
    \centering
    \includegraphics[width=0.8\linewidth]{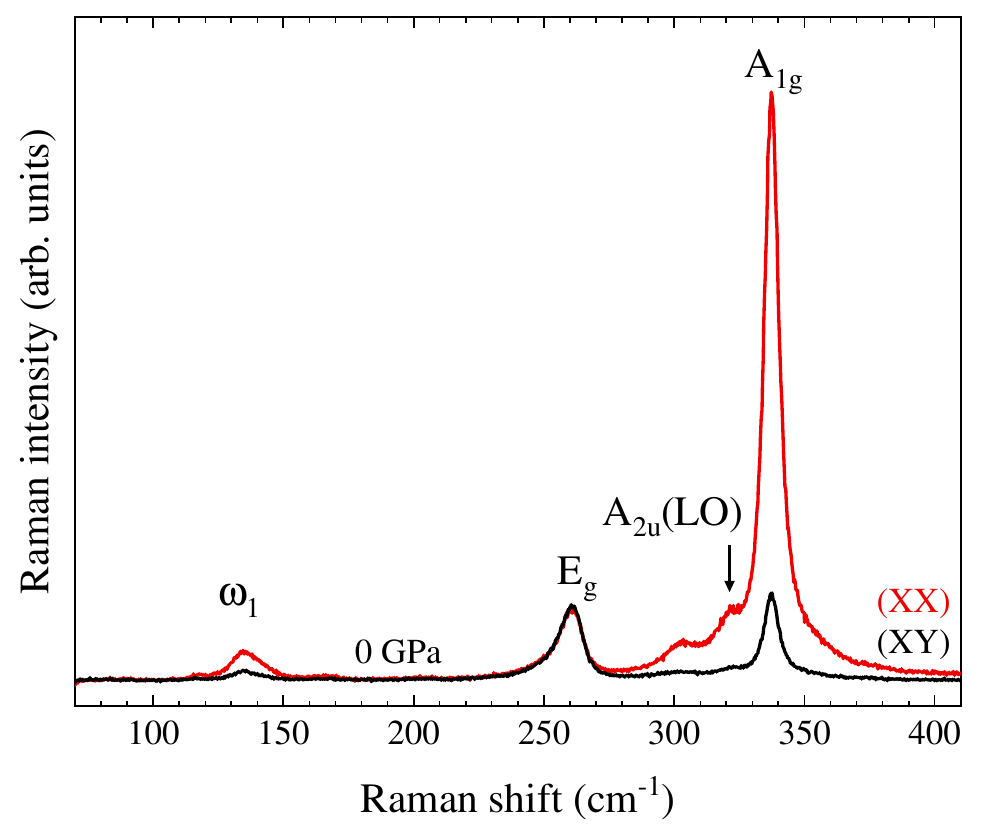}
    \caption
    {\label{fig:561nm_CoCross}
    Raman scattering spectra of \HfS measured in the XX (red) and XY (black) configurations at ambient pressure. 
    The measurements were taken under wavelength $\lambdaup$=632.8~nm excitation with excitation power of 500~$\muup$W.
    }
\end{figure}

\noindent Only the even modes A$_\textrm{1g}$ and E$_\textrm{g}$ are Raman-active.
Important for our analysis is the polarization dependence of the phonon-mode intensities.
The probed co- (XX) and cross-linear (XY) configurations correspond to the polarization of the scattered light with respect to the polarization of the incident laser.
The angle coordinates describe the in-plane spatial orientation of the laser polarization with respect to the \HfS layers.
The polarized RS of \HfS with respect to the XX and XY components can be appreciated in Fig.~\ref{fig:561nm_CoCross}.
In addition to the even modes, two more features are clearly seen,~\cite{iwasaki1982Raman, cingolani1987raman, roubi1988resonance, ibanez2018high} whose phonon dispersion allows us to propose their attribution as 2TA(M) ($\omegaup_1$), and A$_\textrm{2u}$(LO) modes.
The details of the polarization- and ARPRS setup can be found in Methods and in Fig. S1 in the supplementary material (SI).

\begin{figure}[h]
    \centering
    \includegraphics[width=0.9\linewidth]{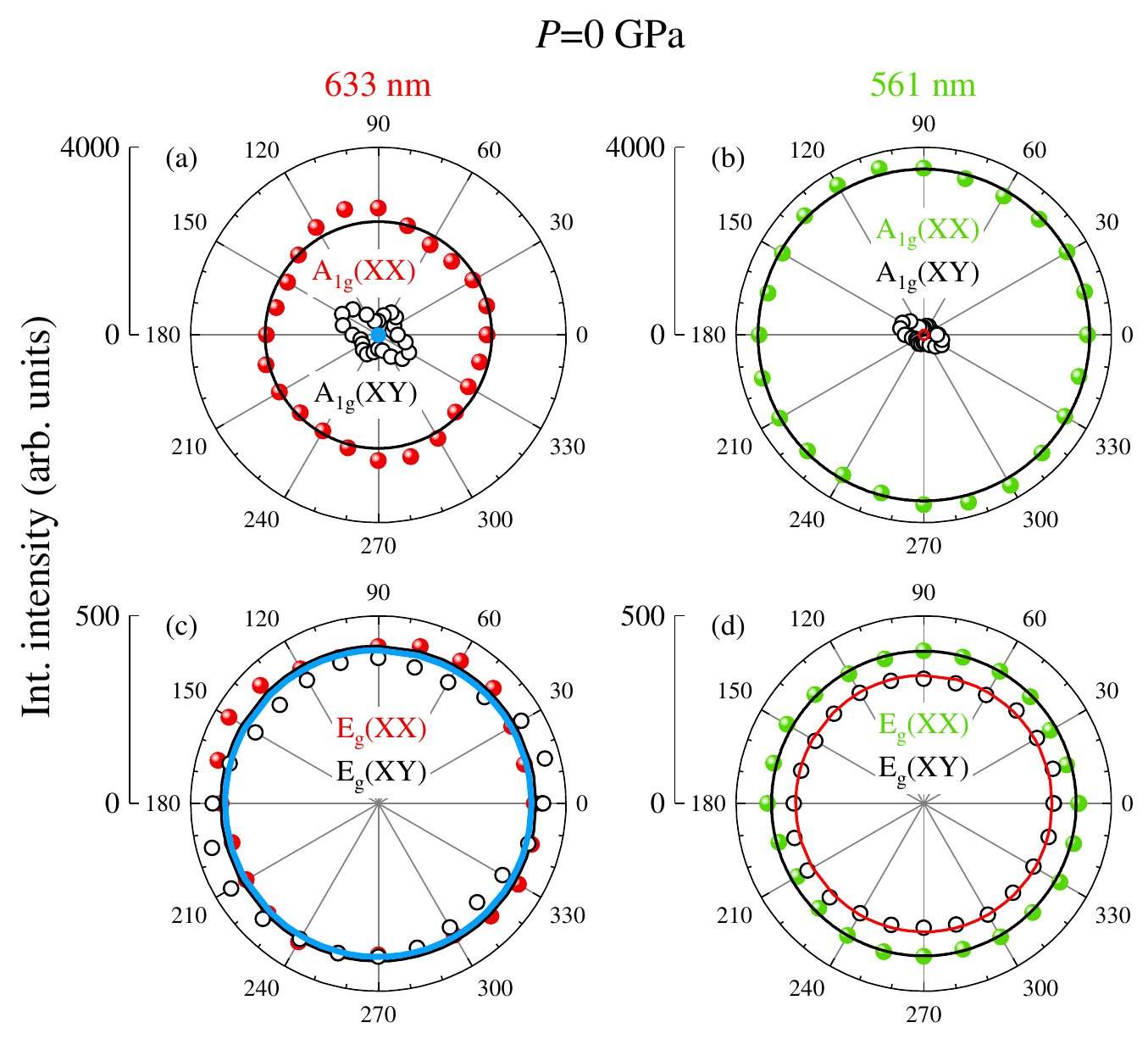}
    \caption
    {\label{fig:RP_pol}
     The ARPRS plots for the A$_\textrm{1g}$ (a, b) and E$_\textrm{g}$ (c, d) modes in both XX (colorful) and XY (black) configuration at ambient pressure.
     The measurements were taken with the $\lambdaup$=633~nm and $\lambdaup$=561~nm excitations.}
\end{figure}

The ARPRS considering A$_\textrm{1g}$ and E$_\textrm{g}$ intensities at ambient pressure is shown in Fig.~\ref{fig:RP_pol}.
The colorful spheres and open circles represent the XX and XY configurations, respectively.
The measurements taken with $\lambdaup$=633~nm and $\lambdaup$=561~nm excitation show that the A$_\textrm{1g}$ intensity does not depend on the polarization direction for the XX configuration and practically vanishes for the XY configuration (although some signal can still be observed, which might be due to light traveling slightly out of the perpendicular direction). 
On the contrary, the E$_\textrm{g}$ intensity in both XX and XY practically does not depend on the polarization of the incoming light for both excitation wavelengths.
The solid lines represent the fittings obtained with the Raman tensors of the 1T-\HfS structure (see Eq.~S1 in the SM).

\subsection{Pressure-dependent Raman scattering} \label{ss:Pdep_RTemp}

\begin{figure}[h]
    \centering
    \includegraphics[width=.85\linewidth]{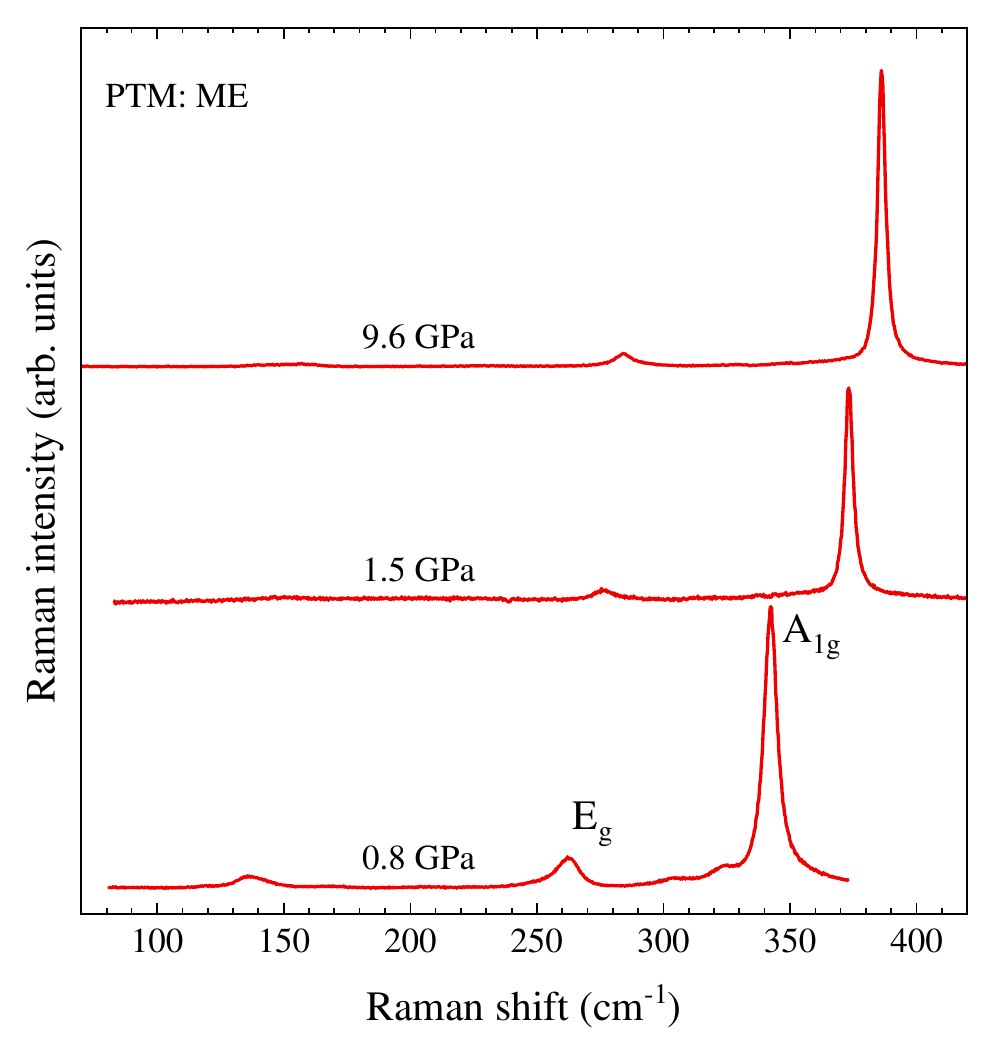}
    \caption
    {\label{fig:Hydrostatic}
     The evolution of the Raman scattering spectra in \HfS during the compression with ME PTM.
     The measurements were taken under $\lambdaup$=633~nm excitation.}
\end{figure}

The evolution of the RS spectra with the pressure transmitted by the ME mixture, which generally provides hydrostatic conditions up to approx. $P$=10~GPa~\cite{Klotz_2009}, is presented in Fig.~\ref{fig:Hydrostatic}.
As can be seen in the figure, the lineshapes of the RS spectra do not change significantly with an increase in pressure up to $P$=9.6 GPa with a systematic blueshift of the observed RS peaks.
In particular, the A$_\textrm{1g}$ peak energy is characterized by the linear dependence in the whole range of the applied pressures (see Fig. S2 in the SM).

\begin{figure*}[ht]
    \centering
    \includegraphics[width=.6\linewidth]{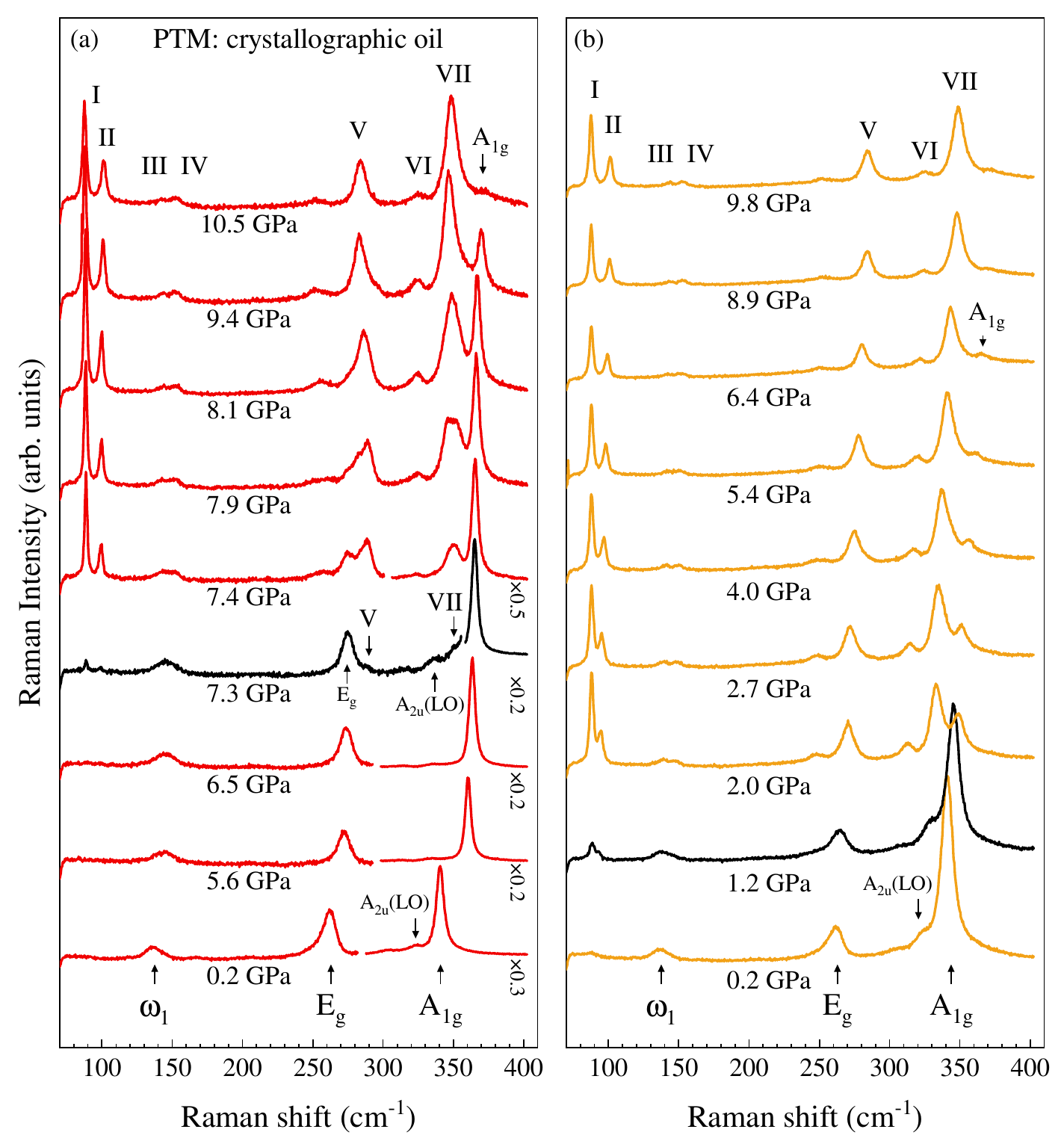}
    \caption
    {\label{RT_PressDep}
    The evolution of the Raman scattering spectra in \HfS during the compression (a) and decompression cycle (b) with the crystallographic oil as PTM.
    The measurements were taken under $\lambdaup$=633~nm excitation.}
\end{figure*}

The RS spectrum under $P$=0.2~GPa, in the experiment with crystallographic oil PTM, is similar to that obtained under ambient conditions (compare Fig.~\ref{RT_PressDep} (a) with Fig.~\ref{fig:561nm_CoCross}).
There are four peaks observed at 136.6~cm$^{-1}$, 262.0~cm$^{-1}$, 323.7~cm$^{-1}$, and 341.1~cm$^{-1}$ referred to as $\omegaup_1$, E$_\textrm{g}$, A$_\textrm{2u}$(LO), and A$_\textrm{1g}$, respectively, in agreement with our previous report.~\cite{Grzeszczyk2022}
The modes will be described below as low-pressure (LP) modes.

The evolution of the RS spectra during compression with crystallographic oil up to $P$=10.5~GPa is shown in Fig.~\ref{RT_PressDep}(a).
The effect of increasing pressure up to $P$=6.5 GPa mainly reveals the blueshifts of the LP modes and the slight changes in their relative intensities.
On the contrary, a substantial change in the RS lineshape is observed at $P$=7.4~GPa, with two well-resolved peaks: I (89~cm$^{-1}$) and II (100~cm$^{-1}$) emerging in the low-frequency range of the spectrum.
Modification of the RS spectra is followed by the emergence of a series of peaks: III (142 cm$^{-1}$), IV (151~cm$^{-1}$), V (290.8~cm $^{-1}$), VI (323.6~cm$^{-1}$), and VII (352.2~cm$^{-1}$).
The I-VII peaks will be referred to as HP modes in the following.
It can be clearly seen in Fig.~\ref{RT_PressDep}(a) that under pressure higher than $P$=7.4~GPa all HP peaks substantially gain intensity compared to their LP counterparts.
With increasing pressure, the A$_\textrm{1g}$ and E$_\textrm{g}$ modes continuously vanishes from the spectra, being barely observed at $P$=10.5~GPa and $P$=8.1~GPa respectively.
A similar evolution of the RS spectra can be observed under the $\lambdaup$=561~nm excitation (see Fig. S3 in the SM).

To probe the reversibility of \HfS structural transformation under pressure, the RS spectra are also recorded during the decompression cycle.
The corresponding selected RS spectra are displayed in Fig.~\ref{RT_PressDep}(b).
The HP RS modes can be observed in the spectra under pressure as low as $P$=1.2~GPa (see black line in Fig.~\ref{RT_PressDep}(b)) at which the A$_\textrm{1g}$ and E$_\textrm{g}$ recover.
Although A$_\textrm{1g}$ emerges as a separate feature at $\sim$350~cm$^{-1}$, the quenching of the V peak accompanies the reemergence of the E$_\textrm{g}$ feature as manifested by a change in the broadening of the peak at $\sim$270~cm$^{-1}$.

\begin{figure}[ht]
    \centering
    \includegraphics[width=.75\linewidth]{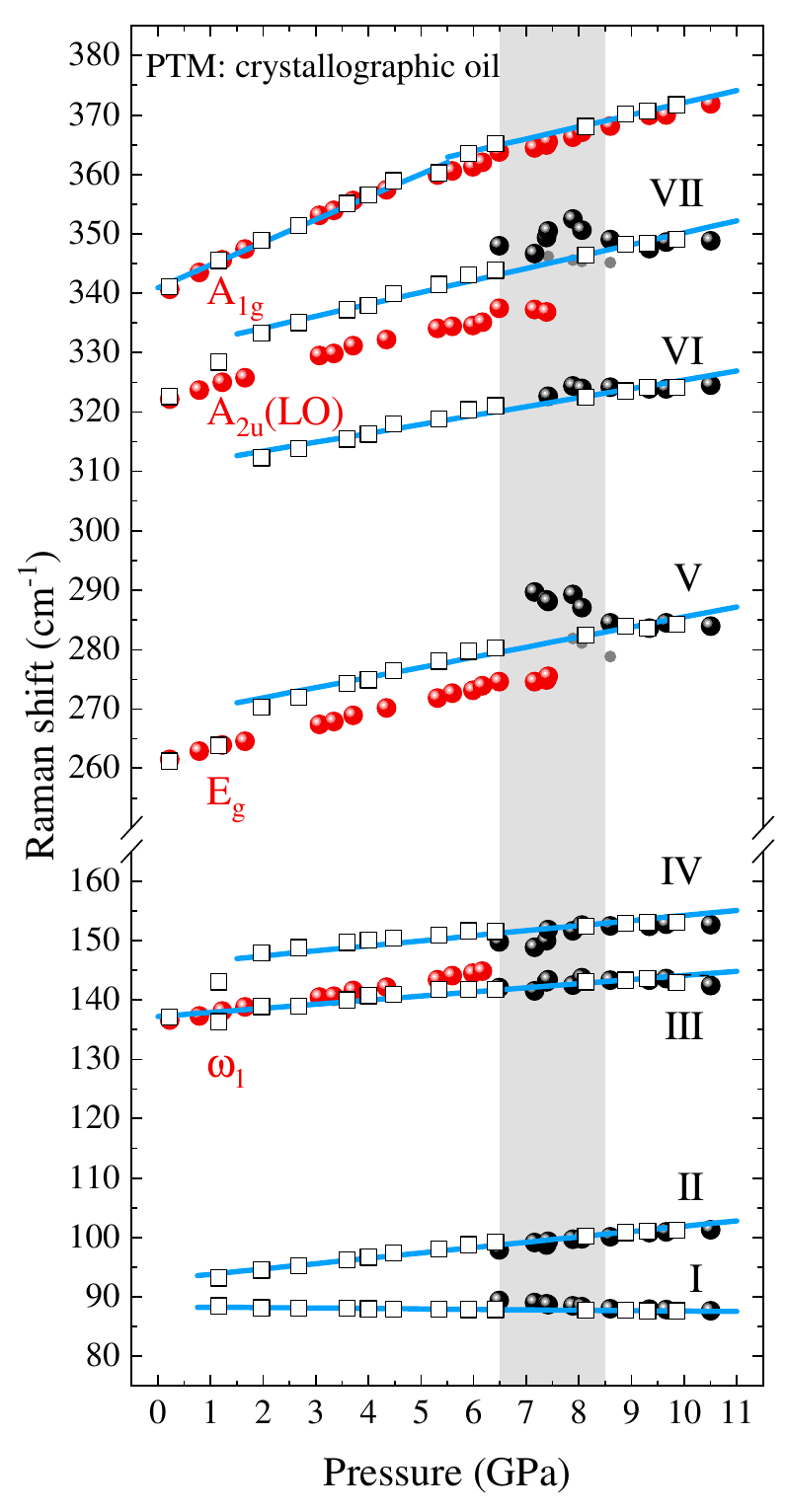}
    \caption
    {\label{Pdep_Rtemp2}
     The pressure evolutions of the Raman shifts obtained for the observed phonon modes. 
     The colorful spheres and open squares represent the results obtained for the compression and decompression processes, respectively.
     The solid blue lines represents the linear fittings to data collected during the decompression cycle.
     The measurements were taken at room temperature under $\lambdaup$=633~nm excitation.
    }
\end{figure}

The energies of the RS peaks measured under the compression (full circles) and decompression (open squares) cycles are shown in Fig.~\ref{Pdep_Rtemp2}. 
An inspection of the pressure evolution of the A$_\textrm{1g}$ energy reveals a slope change at approx. $P$=5~GPa.
An apparently non-monotonic evolution of peaks V and VII is observed at 280~cm$^{-1}$ and~340~cm$^{-1}$) in the pressure range 7.3~GPa<$P$<8.9~GPa (for the determination of their energies, see Fig. S4 in the SM).
This behavior cannot be observed during decompression, as the energies of both peaks follow the monotonic dependence down to $P$=1.2~GPa.
In contrast, the evolution of I, II, IV, and VI during decompression reflect those of the compression procedure down to $P$=1.2~GPa.

Except for the non-monotonic behavior of peaks VII and V, the energies of the Raman modes can be approximated with a linear dependence:
\vspace{-10 pt}
\begin{center}
\begin{equation}
E(P)=E(0)+\alpha\cdot P,
\end{equation}
\end{center}
where $E(0)$ represents the zero-field frequency and $\alpha$ is the pressure coefficient.
The fitting parameters for the observed modes are summarized in Table~\ref{Tbs: RT_HPparameters} regarding the decompression cycle.
The compression cycle is present in the Table SI in the SM.
Note that for the phonon energies, which experience an inflection, the pressure range is given in brackets.

\begin{table}[]
\caption
    {
    Pressure-dependent parameters of phonon modes calculated for decompression cycle of the \HfS: experimental zero-field frequencies - $E(0)$ and their pressure coefficients $\alpha$.
    The results obtained for the measurements taken under $\lambdaup$=633 nm excitation.
    The pressure ranges in which fittings were performed are given in brackets.
    }
\centering
\renewcommand{\arraystretch}{1}
\begin{tabular}[t]{c|c|c}
      & $E(0)$ (cm$^{-1}$) & $\alpha$ (cm$^{-1}$/GPa) \\
      \hline \hline
      $\omegaup_1$  & - & - \\
      E$_\textrm{g}$  & - & - \\
      A$_\textrm{2u}$(LO) & - & - \\
      A$_\textrm{1g}$ ($P$<5.5 GPa) & 340.9 & 3.84\\ 
      A$_\textrm{1g}$ ($P$>5.5 GPa) & 351.7 & 2.04 \\
      I & 88.3 & -0.07 \\
      II & 92.9 & 0.90 \\
      III & 137.2 & 0.69 \\
      IV ($P$>2.0 GPa)& 145.7 & 0.85 \\
      V ($P$>2.0 GPa)& 268.5 & 1.70 \\
      VI ($P$>2.0 GPa)& 310.4 & 1.50 \\
      VII ($P$>2.0 GPa) & 330.1 & 2.01 \\
      \end{tabular}

\label{Tbs: RT_HPparameters}
\end{table}


\subsection{Polarization-resolved Raman scattering in \HfS under high pressure}

\begin{figure}[ht!]
    \centering
    \includegraphics[width=1\linewidth]{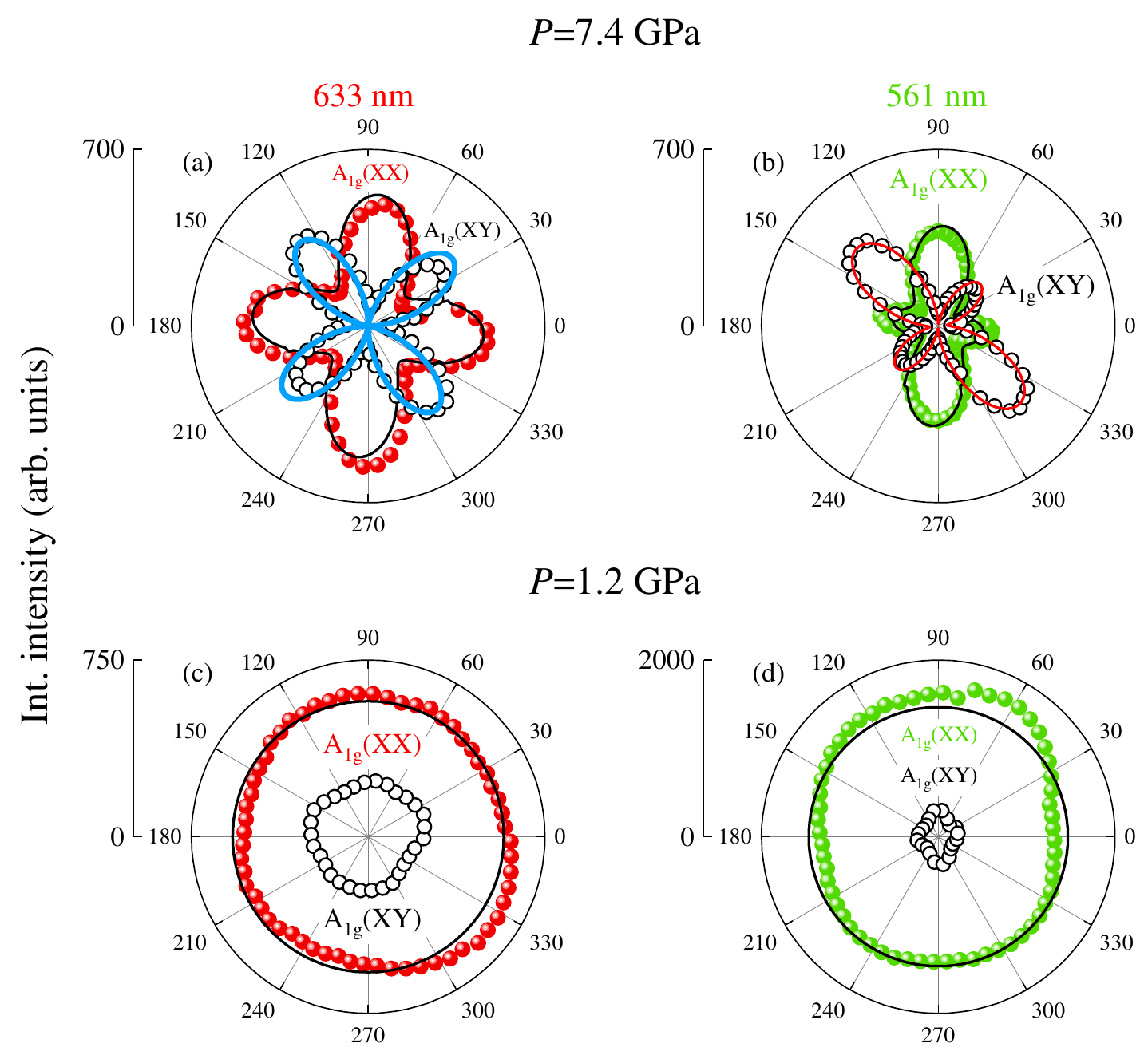}
    \caption
    {\label{fig:HP_A1gPol}
       The ARPRS plots for the A$_\textrm{1g}$ mode in both the XX (red and green full circles) and XY (black open circles) configurations under the $\lambdaup$=633~nm (a, c) and $\lambdaup$=561~nm (b, d) excitation detected upon compression at $P$=7.4~GPa (upper row) and decompression at $P$=1.2~GPa (lower row).
       }
\end{figure}

To gain more insight into the \HfS structure under pressure in crystallographic oil PTM, the polarization-related properties of the observed RS modes were investigated under $P$=7.4~GPa during the compression cycle and at $P$=1.2 GPa during the decompression cycle.
Let us first address the polarization dependence of the A$_\textrm{1g}$ intensity, which is still present in the RS spectrum at $P$=7.4~GPa (see Fig.~\ref{fig:HP_A1gPol})
A four-fold symmetry of the A$_\textrm{1g}$ intensity is observed for both the XX and XY configurations under the $\lambdaup$=633 nm excitation.
In contrast, a two-fold symmetry of the A$_\textrm{1g}$ intensity is seen for both the XX and XY configurations under the $\lambdaup$=561 nm excitation.
For both excitation energies, the main axis of the A$_\textrm{1g}$ intensity for the XY configuration is rotated by 45$^{\circ}$ with respect to the main axis for the XX configuration.
Note that only small deviations from the circular dependence of the A$_\textrm{1g}$ intensity can be appreciated for both the $\lambdaup$=633~nm and $\lambdaup$=561~nm excitations measured at $P$=1.2 GPa.
Showing that 1t-\HfS phase can be recovered.
For the XY configuration, the A$_\textrm{1g}$ mode has a much lower intensity with a negligible angle dependence.
The ARPRS concerning the rest of observed modes under $P$=1.2 GPa are present in Fig. S5 in the SM.
The solid lines represents the fittings.
Under $P$=7.4 GPa we used the Raman tensors from Eq.~\ref{Tb: RT_Tensor1} and Eq.~\ref{Tb: RT_Tensor3} for $\lambdaup$=633~nm and $\lambdaup$=561~nm  excitation, respectively. Meanwhile, the ambient 1T-\HfS tensors (see Eq.~S1 in the SM) were considered to $P$=1.2 GPa.
The ARPRS intensities of the HP modes depend on the excitation energy.
The detailed analysis of the presented results is described in the following.

\begin{figure}[]
    \centering
    \includegraphics[width=.95\linewidth]{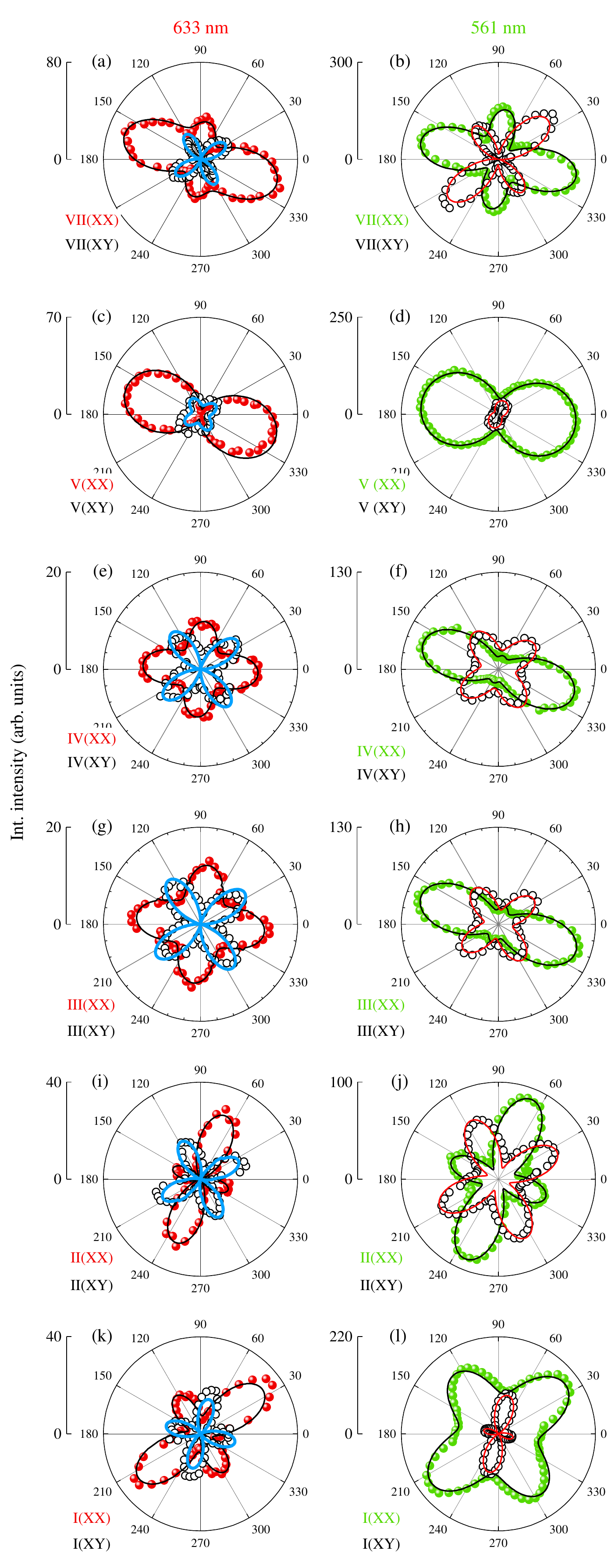}
    \caption
    {\label{fig:HPpol1}
    The polar plots for the HP Raman scattering modes: VII (a, b), V (c, d)), IV (e, f), III (g, h), II (i, j), and I (k, l) under the $\lambdaup$=632.8~nm (left-hand side column) and $\lambdaup$=561~nm (right-hand side column) excitations detected upon compression at $P$=7.4 GPa.
    The full and open circles represent the XX and XY configuration, respectively.
    }
\end{figure}

\section{Discussion} \label{ss:Pdep_RTemp2}

The phonon intensity in the RS experiment calculated by the Placzek approximation can be expressed as~\cite{Porezag1996, Carvalho2015, steele2016}:

\vspace{-10 pt}
\begin{center}
\begin{equation}
I=I_0 \cdot \sum_{j}^{}|\hat{e}_i \cdot R_{j} \cdot \hat{e}_s|^2,
\end{equation}
\end{center}

\noindent where $I_0$ reflects experimental parameters such as laser intensity, integration time, $etc.$
The $\hat{e}_i$ and $\hat{e}_s$ are the electric field vectors of the incident and scattered light, respectively, and $R$ represents the Raman tensor for different symmetry modes.

Let us restrict our discussion to light traveling along the z axis, perpendicular to the crystal planes, and adopt the counterclockwise chirality in our setup.
The XX configuration in this case sets the electric field of the incident light ($\hat{e}_i$) and the scattered beam ($\hat{e}_s$) as $\hat{e}_i$=$\hat{e}_s$=(cos$\thetaup$, sin$\thetaup$).
The XY configuration is reached by rotating the detection polarizer by 90$^\circ$, thus $\hat{e}_s$=($-$sin$\thetaup$, cos$\thetaup$).

The 1T-\HfS is a hexagonal layered crystal with a three-fold symmetry perpendicular to the layer planes.
Its RS spectrum is isotropic; it does not depend on the polarization of light traveling perpendicularly to the plane.
The Raman tensor in the case of a symmetric $\textrm{A}_\textrm{1g}$ mode of \HfS under ambient conditions is diagonal with equal matrix elements and reads:

\vspace{-10 pt}
\begin{equation}
\textrm{R}=\begin{pmatrix} a & 0\\ 0 & a\end{pmatrix} 
\label{Tb: TensorA1}
\end{equation}
\noindent where $a$ represents the absolute values of the tensor elements.
This leads to a symmetric polar dependence of the RS in the XX configuration and to the quenching of the mode in the XY configuration, which corresponds to the results obtained under ambient conditions for A$_\textrm{1g}$ (see Fig.~\ref{fig:RP_pol}).

More complicated is the polar dependence of the RS feature which evolved from the "ambient condition" A$_\textrm{1g}$ mode at 7.4 GPa (Fig.~\ref{fig:HP_A1gPol} (a, b)).
Let us start our discussion with the results obtained with $\lambdaup$=633 nm excitation.
As can be seen in Fig.~\ref{fig:HP_A1gPol} (a), the polar dependence of the A$_\textrm{1g}$ mode for both XX and XY configurations exhibit a four-fold symmetry.
This clearly shows that it cannot be described by the Raman tensor presented in Eq.~\ref{Tb: TensorA1}.
To account for the results presented in Fig.~\ref{fig:HP_A1gPol}, we consider the more general form of the Raman tensor~\cite{Cardona1982}, which corresponds to A/A$_1$/A$_\textrm{g}$ Raman-active modes in a crystal of the orthorhombic symmetry\cite{Pimenta2021} and can be described as:


\begin{equation}
\textrm{R}=\begin{pmatrix} ae^{i\varphi_a} & 0  \\ 0 & be^{i\varphi_b} \end{pmatrix} = 
ae^{i\varphi_a}\begin{pmatrix} 1 & 0\\ 0 & (b/a)e^{i\varphi_{ba}} \end{pmatrix}
\label{Tb: RT_Tensor1}
\end{equation}
\noindent from now on, the lowercase letter will always represents the absolute values of the Raman tensor elements and $\varphi_i$ the related complex phases due to the absorption effects.
Note that the complex nature of the Raman tensor elements is proposed to reflect the optical anisotropy of a crystal.~\cite{kranert2016}
Following the discussion presented in Ref.~\cite{Pimenta2021}, we note that the only relative values of $b$/$a$ and $\varphi_{ba}$=$\varphi_b$-$\varphi_a$ can be determined from the experimental results.
The polar dependence of the RS intensity expected for the Raman tensor for A/A$_1$/A$_\textrm{g}$ modes in a crystal of orthorhombic symmetry in the XX and XY configurations can be found in Eq.~S9-11 in the SM.
As is seen in Fig.~\ref{fig:HP_A1gPol} (a), the results of the fit with $b$/$a$=0.94 and $\varphi_{ba}$=103$^{\circ}$ correspond very well to the experimental results.
In our opinion, this strongly suggests that the symmetry of the \HfS crystal changes upon the compression cycle.
Due to the observed results for two different PTM, we assume that the modification of the \HfS symmetry is related to the non-hydrostacity provided by the crystallographic oil.
The non-hydrostatic component of pressure is confirmed by increasing splitting between the R$_1$ and R$_2$ ruby-related luminescence lines (see Fig. S2 (b) in the SM).
The component affects the symmetry of the crystal and the observed Raman-active out-of-plane vibrations, and the crystal structure becomes orthorhombic (with three crystallographic axes perpendicular to each other that have unequal length).
The principal Raman active out-of-plane vibration of the A$_\textrm{1g}$ symmetry in the 1T-\HfS crystal under ambient conditions becomes A/A$_1$/A$_\textrm{g}$ vibration, depending on the actual distortion of the crystal.
However, for the sake of clarity, we will refer to the mode as A$_\textrm{1g}$ also above the transformation.
It should also be noted that the change in crystal symmetry probably also affects the pressure coefficient for the A$_\textrm{1g}$ mode above the transformation.
As can be seen in Fig.~\ref{RT_PressDep}, the pressure coefficient changes at approx. $P$=5~GPa from 3.93 cm$^{-1}$/GPa to 2.38 cm$^{-1}$/GPa (see Table~\ref{Tbs: RT_HPparameters}).
This kind of change is characteristic only for non-hydrostatic pressure conditions (see Fig. S2 (a) in the SM).

We can appreciate the ARPRS to HP modes in Fig.~\ref{fig:HPpol1} for $\lambdaup$=633~nm (left column) and $\lambdaup$=561~nm (right column).
Focusing on $\lambdaup$=633~nm results, the polar dependence for the vibrations obtained from Eq.~\ref{Tb: RT_Tensor1} shows a four-fold symmetry in the XY configuration, which is the case for II (i), III (g), and IV (e) (fitting parameters can be found in Table SII in the SM).
The rest modes show a more trick dependence, and to account for them, one must consider the Raman tensor, which corresponds to the A$_\textrm{g}$ mode in a triclinic crystal,~\cite{Pimenta2021} which can be given as:

\vspace{-10 pt}
\begin{equation}
\small
\textrm{R}=\begin{pmatrix} ae^{i\varphi_a} & de^{i\varphi_d}  \\ de^{i\varphi_d} & be^{i\varphi_b} \end{pmatrix}
= ae^{i\varphi_a}\begin{pmatrix}1 & (d/a)e^{i\varphi_{da}}\\ (d/a)e^{i\varphi_{da}} & (b/a)e^{i\varphi_{ba}} \end{pmatrix}
\label{Tb: RT_Tensor2}
\end{equation}

\noindent As previously, we note that only the relative values can be determined from the experimental results, $e.g.$ $b$/$a$ and $\varphi_{ba}$=$\varphi_b$-$\varphi_a$.
In Fig.~\ref{fig:HPpol1} the Raman tensor in Eq.~\ref{Tb: RT_Tensor2} describes well the modes I (k), V (c), and VII (a).
The parameters and the polar dependence of the RS to XX and XY configurations can be found in Eq. S12-14 in SM.

Analyzing our observations, we first address the emergence of the anisotropic components of the Raman tensor.
In our opinion, they must result from the anisotropy induced by the PTM used in our experiment.
The crystallographic oil is known to introduce the non-hydrostatic component to the pressure.
The anisotropy is removed during the decompression cycle at about $P$=1.2~GPa, as clearly seen in Fig.~\ref{fig:HP_A1gPol} (c, d).
The use of crystallographic oil in pressure experiments strongly affects the pressure value at which the phase transition occurs.
We also found that with the use of the ME, the phase transition is not apparent up to $P$=9.6 GPa (Fig.~\ref{fig:Hydrostatic}).
The non-hydrostatic component of pressure, clearly leads to distortion of the 1T-\HfS crystallographic structure as observed by the anisotropic RS spectrum of the A$_\textrm{1g}$ mode obtained at $P$=7.4~GPa under the $\lambdaup$=633 nm excitation, see Fig.~\ref{fig:HP_A1gPol} (a).
The Raman tensor, which can be used to interpret the results, suggests the orthorhombic crystal structure of the distorted \HfS. 
Similar anisotropy has been reported in a series of different layered materials, $e.g.$ WTe$_2$~\cite{Song2016}, black phosphorus~\cite{Ribeiro2015}, black arsenic~\cite{Wang2021}, $\alpha$ MoO$_3$~\cite{Gong2022}, ReS$_2$~\cite{Chenet2015} or GeS~\cite{Zawadzka2021}.

Some of the actual Raman tensors, which describe the polar dependence of the HP peaks (II, III, and IV) and (I, V, and VII), correspond to the orthorhombic and triclinic crystal structures, respectively. 
We cannot offer a firm explanation for this observation.
However, the energy coincidence of the III and $\omegaup_1$ peaks may suggest that peaks III and IV are split features, which evolve from the $\omegaup_1$ vibration in the unaffected \HfS crystal.
Similarly, the peak II may be due to other vibrations, which are not optically active in the original \HfS crystal and become visible in the distorted structure. 
This would refer to the peaks I, V, VII (and most likely VI, whose intensity was too low to analyze properly) to a distorted orthorhombic symmetry of the HP phase of \HfS.

Let us focus on the orthorhombic $Pnma$ phase that was recently proposed as an HP phase of \HfS.~\cite{Zhang_2022}
There are 18 Raman-active modes in the $Pnma$ phase of \HfS: 
\vspace{-20 pt}
\begin{center}
\begin{equation}
\Gamma=6\textrm{A}_\textrm{g} + 3\textrm{B}_\textrm{1g}+ 6\textrm{B}_\textrm{2g}+ 3\textrm{B}_\textrm{3g}
\end{equation}
\end{center}

\begin{figure}[ht!]
    \centering
    \includegraphics[width=1\linewidth]{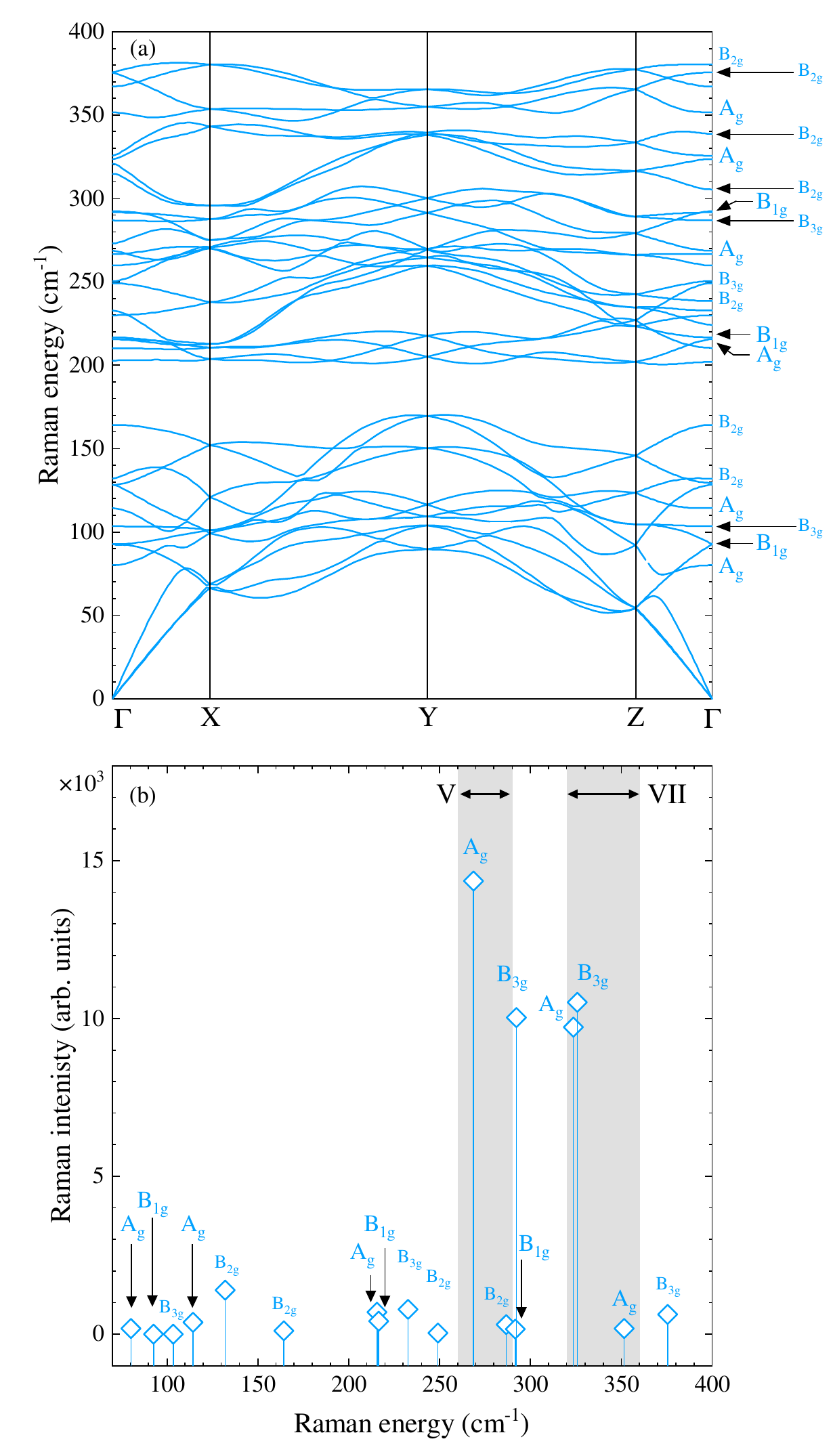}
    \caption
    {\label{fig:Prezentacja1}
       (a) The phonon dispersion for the orthorhombic \HfS $Pnma$ phase calculated using LDA functional and (b) the corresponding relative intensities of the RS peaks.
       The gray stripes indicate the energy range, where the peaks V and VII were observed in the experiment.
       }  
\end{figure}

Only $\textrm{A}_\textrm{g}$ and $\textrm{B}_\textrm{1g}$ are visible in the back-scattering experiment.
Therefore, nine Raman-active modes for the $Pnma$ phase are expected in the measured HP RS spectrum.
Using density functional theory (DFT), we calculated the phonon dispersion in $Pnma$ \HfS, see Fig.~\ref{fig:Prezentacja1}(a).
To better visualize the Raman-active modes at the $\Gamma$ point of the Brillouin zone (BZ), we determined the energies of the modes and their relative intensities, as shown in Figure~\ref{fig:Prezentacja1}.
Quantitative correspondence with the experimental results cannot be expected, as the calculations are made for the ambient pressure at $T$=0~K.
Moreover, the LDA functional is used, which is known to underestimate the lattice constants that affect the frequencies and intensities of the modes.
Possible resonant conditions are also neglected.
There are, however, two strong $\textrm{A}_\textrm{g}$ modes with the energies at 269~cm$^{-1}$ and 324~cm$^{-1}$, while other $\textrm{A}_\textrm{g}$ and $\textrm{B}_\textrm{1g}$ modes are much weaker.
As can be seen in Table~\ref{Tbs: RT_HPparameters}, their energies coincide with the extrapolated zero-pressure energies of peaks V and VII.
We note that the polarization-resolved dependence of the intensities of peaks V and VII (as well as peak I) correspond to the triclinic crystal symmetry, which may result from the pressure-induced distortion of the orthorhombic structure.

Finally, let us address the polar dependence of the peaks measured with $\lambdaup$=561 nm excitation.
To describe the polar dependence present in Fig.~\ref{fig:HPpol1} of the II (j), III (h), and IV (f) intensities obtained under the $\lambdaup$=561 nm excitation, the Raman tensor shown in Eq.~\ref{Tb: RT_Tensor2} (symmetric with non-diagonal components) should be applied. 
We note the apparent difference between that observation and the results obtained with the $\lambdaup$=633~nm excitation, which can be fitted using the Raman tensor shown in Eq~\ref{Tb: RT_Tensor1}.
The strong effect of the excitation energy on the polar dependence of RS peaks is usually attributed to a possible symmetry-dependent resonant exciton-phonon interaction.~\cite{Lorchat2016}
Furthermore, it was demonstrated that the intensity of RS for one of the A$_\textrm{1g}$ peaks in 3 layer MoTe$_2$ strongly depends on the excitation energy.~\cite{Froehlicher}
The total intensity of the RS was shown to be derived from the contributions from different points of the BZ.
As positive and negative contributions enter the final result (to be squared), the RS intensity can substantially vary as a function of excitation energy, leading, for example, to complete quenching of the RS intensity (as at $T$=200~K for 2L MoTe$_2$~\cite{grzeszczyk2016Raman})
We propose that a similar effect of the electric field direction can be observed here, however, a more strict theoretical approach is needed here, which is beyond the scope of this work.

The most striking are the angle-dependent evolution of the $\textrm{A}_\textrm{1g}$ and HP peaks VII, V, and I, presented in Fig.~\ref{fig:HP_A1gPol} (b) and Fig.~\ref{fig:HPpol1} (b, d, and l), respectively.
To interpret these experimental observations, a most general non-symmetric form of a tensor must be assumed with unequal off-diagonal elements, which reads:

\vspace{-10 pt}
\begin{equation}
\small
\textrm{R}=\begin{pmatrix} 
ae^{i\varphi_a} & fe^{i\varphi_d}  \\ 
ce^{i\varphi_c} & be^{i\varphi_b} 
\end{pmatrix}
= ae^{i\varphi_a}\begin{pmatrix}
1 & (f/a)e^{i\varphi_{da}}\\ 
(c/a)e^{i\varphi_{ca}} & (b/a)e^{i\varphi_{ba}} 
\end{pmatrix}
\label{Tb: RT_Tensor3}
\end{equation}

\noindent The tensor in Eq.~\ref{Tb: RT_Tensor3} produces accurate fittings to the above mentioned modes.
The calculated parameters, as well as, polar dependence of the RS to XX and XY configurations are present in Table SII and Eq.~S16-17 in the SM.
Tensors like such one were previously reported to explain the RS in 1T-TaS$_2$~\cite{Lacinska2022} or PrCl$_3$~\cite{Koninstein1969}.
However, more strict theoretical analysis is needed to understand our results.
In the case of peak I, the complex phases $\varphiup_\textrm{ca}$ and $\varphiup_\textrm{fa}$ are anti-symmetric (180°), but the same it is not true for the others.
As under $\lambdaup$=633~nm we can describe our data with conventional symmetric tensors (Eq.~\ref{Tb: RT_Tensor1} and Eq.~\ref{Tb: RT_Tensor2}), we believe resonant effects are taking place under $\lambdaup$=561~nm breaking the tensor symmetry.


\section{Conclusions \label{sec:Conclusions}}

The scattering of the bulk \HfS was investigated using the $\lambdaup$=632.8 nm and $\lambdaup$=561 nm excitations at room temperature under external pressure up to $P$=10.5 GP applied under hydrostatic and non-hydrostatic conditions.
No significant RS spectrum lineshape was observed with a systematic blueshift of the RS features under hydrostatic conditions.
A phase transition in \HfS was observed at $P\approx$7~GP when a non-hydrostatic environment was provided.
The phase transition manifested itself with the appearance of seven new modes in the RS spectrum.
The angle- and polarization-dependent RS spectroscopic study performed at $P$=7.4~GPa clearly showed anisotropic behavior for all observed RS modes.
It was proposed that the original ambient-pressure 1T-\HfS phase was distorted under pressure, leading to the orthorhombic crystal structure.
Simultaneously, a new phase appeared, which was related to a distorted $Pnma$ phase.
The pressure-induced phase coexisted with the ambient-condition phase up to $P$=10.5~GPa and persisted during the decompression down to $P$=1.2~GPa confirms its metastable behavior.
The polarization dependence of the RS peaks recorded at 7.4 GPa with $\lambda$=561~nm excitation are more complex than those acquired at $\lambda$=633~nm.
Some modes under green excitation can only be fitted by using a non-symmetric Raman tensor, and this can be attributed to resonant effects.
Therefore, more work regarding the bandgap dependence of the HP phases of \HfS will be necessary in order to confirm this conclusion and the overall interpretation of the present ARPRS experiments.


\section*{Methods\label{sec:Methods}}

The investigated \HfS crystals were obtained from 2D semiconductors.
The iodine-related photoluminescence signal~\cite{Zawadzka2023} was observed in the 2D semiconductor crystal.
However, in our opinion, the presence of iodine has no significant effect in the pressure range investigated.
The X-Ray diffraction measurements found to be in perfect agreement with that of the octahedral 1T phase of \HfS~\cite{lukovsky1973IR}, although a minor amount of an unidentified phase was also detected.~\cite{ibanezPrivate}

Room-temperature optical measurements were made by excitation of the sample with a diode pumped laser at $\lambdaup$=561~nm (2.21 eV) or He-Ne laser at $\lambdaup$=632.8~nm (1.96 eV) focused by a 50$\times$ long-working distance objective with a 0.55 (NA). 
The excitation power focused on the sample was kept at approx. 500~$\mu$W in all measurements.
The scattered light was collected in the back-scattering geometry, sent through a 0.75~m monochromator, and then detected by means of a liquid nitrogen-cooled CCD camera. 

The polarization-sensitive spectroscopy was carried out by introducing a $\lambdaup/2$ plate directly on top of the objective lens to simultaneously rotate the incoming and scattered light with reference to the investigated in-plane crystal orientation.
The experiments were conducted with the incoming and scattered light travelling along $z$-axis.
Both configurations, $i.e.$ co- (XX) and cross-linear (XY), were probed that  correspond to the detection polarizer aligned parallel or perpendicular to the excitation one, respectively.
More details and the schematic illustration of the experimental setup can be found in the Fig. S1 in the SM.

The high-pressure experiments were conducted at room temperature in a non-magnetic diamond anvil cell (SymmDAC 60 LT DAC) made of BeCu alloy~\cite{Graf2011}.
The pressure chamber was sealed using a BeCu gasket with a height of approx. 75 $\muup$m, a diameter of 250 $\muup$m, and filled with a ME or crystallographic oil.
The pressure was monitored by the ruby  (Al$_2$O$_3$:Cr$^{+3}$) fluorescent method.
The $R_1-R_2$-separation is used as a measure of uniaxial stress~\cite{Takemura2007}.

Density functional theory (DFT) calculations were performed in Quantum Espresso \cite{Giannozzi_2009,Giannozzi_2017} with scalar-relativistic norm-conserving pseudopotentials cite{PhysRevB.88.085117}. Local density approximation (LDA) to the exchange-correlation functional was used. The wavefunction and charge density cutoff energies were set to 100 and 400 Ry, respectively. A 6$\times$10$\times$4 $\Gamma$-centered Monkhorst-Pack $k$-points grid was used. Geometrical structure was fully optimized with 10$^{-5}$ Ry/Bohr and 0.01 kbar criteria for the interatomic forces and stress tensor components, respectively. Phonon and non-resonant Raman calculations were performed within Density Functional Perturbation Theory. A 2$\times$2$\times$2 $q$-points grid was found sufficient to converge phonon dispersion and Raman intensities. 

\section*{Supplementary Material}
In the supplementary material is present the schematic illustration of the experimental setup, fundamental pressure features, more results considering $\lambdaup$=561~nm excitation, details related to fittingss and parameters tables.

\section*{Data availability statement}
The data that support the findings of this study are available from the corresponding author upon reasonable request.

\section*{Acknowledgments}
The work has been supported by the Excellence Initiative - Research University at the University of Warsaw. 
Z.M. and W.Z. acknowledge support from the National Natural Science Foundation of China (grant no. 62150410438), the International Collaboration Project (no. B16001), and the Beihang Hefei Innovation Research Institute (project no. BHKX-19-02).
This research was carried out with the support of the Interdisciplinary Centre for Mathematical and Computational Modelling University of Warsaw (ICM UW) under computational allocation no G95-1773.

\clearpage
\bibliographystyle{apsrev4-2}
\bibliography{biblio}

\begin{thebibliography}{44}%
\makeatletter
\providecommand \@ifxundefined [1]{%
 \@ifx{#1\undefined}
}%
\providecommand \@ifnum [1]{%
 \ifnum #1\expandafter \@firstoftwo
 \else \expandafter \@secondoftwo
 \fi
}%
\providecommand \@ifx [1]{%
 \ifx #1\expandafter \@firstoftwo
 \else \expandafter \@secondoftwo
 \fi
}%
\providecommand \natexlab [1]{#1}%
\providecommand \enquote  [1]{``#1''}%
\providecommand \bibnamefont  [1]{#1}%
\providecommand \bibfnamefont [1]{#1}%
\providecommand \citenamefont [1]{#1}%
\providecommand \href@noop [0]{\@secondoftwo}%
\providecommand \href [0]{\begingroup \@sanitize@url \@href}%
\providecommand \@href[1]{\@@startlink{#1}\@@href}%
\providecommand \@@href[1]{\endgroup#1\@@endlink}%
\providecommand \@sanitize@url [0]{\catcode `\\12\catcode `\$12\catcode
  `\&12\catcode `\#12\catcode `\^12\catcode `\_12\catcode `\%12\relax}%
\providecommand \@@startlink[1]{}%
\providecommand \@@endlink[0]{}%
\providecommand \url  [0]{\begingroup\@sanitize@url \@url }%
\providecommand \@url [1]{\endgroup\@href {#1}{\urlprefix }}%
\providecommand \urlprefix  [0]{URL }%
\providecommand \Eprint [0]{\href }%
\providecommand \doibase [0]{https://doi.org/}%
\providecommand \selectlanguage [0]{\@gobble}%
\providecommand \bibinfo  [0]{\@secondoftwo}%
\providecommand \bibfield  [0]{\@secondoftwo}%
\providecommand \translation [1]{[#1]}%
\providecommand \BibitemOpen [0]{}%
\providecommand \bibitemStop [0]{}%
\providecommand \bibitemNoStop [0]{.\EOS\space}%
\providecommand \EOS [0]{\spacefactor3000\relax}%
\providecommand \BibitemShut  [1]{\csname bibitem#1\endcsname}%
\let\auto@bib@innerbib\@empty
\bibitem [{\citenamefont {Deng}\ \emph {et~al.}(2018)\citenamefont {Deng},
  \citenamefont {Sumant},\ and\ \citenamefont {Berry}}]{DENG201814}%
  \BibitemOpen
  \bibfield  {author} {\bibinfo {author} {\bibfnamefont {S.}~\bibnamefont
  {Deng}}, \bibinfo {author} {\bibfnamefont {A.~V.}\ \bibnamefont {Sumant}},\
  and\ \bibinfo {author} {\bibfnamefont {V.}~\bibnamefont {Berry}},\ }\href
  {https://doi.org/https://doi.org/10.1016/j.nantod.2018.07.001} {\bibfield
  {journal} {\bibinfo  {journal} {Nano Today}\ }\textbf {\bibinfo {volume}
  {22}},\ \bibinfo {pages} {14} (\bibinfo {year} {2018})}\BibitemShut {NoStop}%
\bibitem [{\citenamefont {Pimenta~Martins}\ \emph {et~al.}(2023)\citenamefont
  {Pimenta~Martins}, \citenamefont {Comin}, \citenamefont {Matos},
  \citenamefont {Mazzoni}, \citenamefont {Neves},\ and\ \citenamefont
  {Yankowitz}}]{Pimenta2023}%
  \BibitemOpen
  \bibfield  {author} {\bibinfo {author} {\bibfnamefont {L.~G.}\ \bibnamefont
  {Pimenta~Martins}}, \bibinfo {author} {\bibfnamefont {R.}~\bibnamefont
  {Comin}}, \bibinfo {author} {\bibfnamefont {M.~J.~S.}\ \bibnamefont {Matos}},
  \bibinfo {author} {\bibfnamefont {M.~S.~C.}\ \bibnamefont {Mazzoni}},
  \bibinfo {author} {\bibfnamefont {B.~R.~A.}\ \bibnamefont {Neves}},\ and\
  \bibinfo {author} {\bibfnamefont {M.}~\bibnamefont {Yankowitz}},\ }\href
  {https://doi.org/10.1063/5.0123283} {\bibfield  {journal} {\bibinfo
  {journal} {Applied Physics Reviews}\ }\textbf {\bibinfo {volume} {9}},\
  \bibinfo {pages} {011313} (\bibinfo {year} {2023})}\BibitemShut {NoStop}%
\bibitem [{\citenamefont {Pei}\ \emph {et~al.}(2022)\citenamefont {Pei},
  \citenamefont {Wang},\ and\ \citenamefont {Xia}}]{PEI2022}%
  \BibitemOpen
  \bibfield  {author} {\bibinfo {author} {\bibfnamefont {S.}~\bibnamefont
  {Pei}}, \bibinfo {author} {\bibfnamefont {Z.}~\bibnamefont {Wang}},\ and\
  \bibinfo {author} {\bibfnamefont {J.}~\bibnamefont {Xia}},\ }\href
  {https://doi.org/https://doi.org/10.1016/j.matdes.2021.110363} {\bibfield
  {journal} {\bibinfo  {journal} {Materials and Design}\ }\textbf {\bibinfo
  {volume} {213}},\ \bibinfo {pages} {110363} (\bibinfo {year}
  {2022})}\BibitemShut {NoStop}%
\bibitem [{\citenamefont {Zhang}\ \emph {et~al.}(2020)\citenamefont {Zhang},
  \citenamefont {Tang}, \citenamefont {Khan}, \citenamefont {Hasan},
  \citenamefont {Wang}, \citenamefont {Yan}, \citenamefont {Yildirim},
  \citenamefont {Torres}, \citenamefont {Neupane}, \citenamefont {Zhang},
  \citenamefont {Li},\ and\ \citenamefont {Lu}}]{Zhang2020}%
  \BibitemOpen
  \bibfield  {author} {\bibinfo {author} {\bibfnamefont {L.}~\bibnamefont
  {Zhang}}, \bibinfo {author} {\bibfnamefont {Y.}~\bibnamefont {Tang}},
  \bibinfo {author} {\bibfnamefont {A.~R.}\ \bibnamefont {Khan}}, \bibinfo
  {author} {\bibfnamefont {M.~M.}\ \bibnamefont {Hasan}}, \bibinfo {author}
  {\bibfnamefont {P.}~\bibnamefont {Wang}}, \bibinfo {author} {\bibfnamefont
  {H.}~\bibnamefont {Yan}}, \bibinfo {author} {\bibfnamefont {T.}~\bibnamefont
  {Yildirim}}, \bibinfo {author} {\bibfnamefont {J.~F.}\ \bibnamefont
  {Torres}}, \bibinfo {author} {\bibfnamefont {G.~P.}\ \bibnamefont {Neupane}},
  \bibinfo {author} {\bibfnamefont {Y.}~\bibnamefont {Zhang}}, \bibinfo
  {author} {\bibfnamefont {Q.}~\bibnamefont {Li}},\ and\ \bibinfo {author}
  {\bibfnamefont {Y.}~\bibnamefont {Lu}},\ }\href
  {https://doi.org/https://doi.org/10.1002/advs.202002697} {\bibfield
  {journal} {\bibinfo  {journal} {Advanced Science}\ }\textbf {\bibinfo
  {volume} {7}},\ \bibinfo {pages} {2002697} (\bibinfo {year}
  {2020})}\BibitemShut {NoStop}%
\bibitem [{\citenamefont {Zhang}\ \emph {et~al.}(2014)\citenamefont {Zhang},
  \citenamefont {Huang}, \citenamefont {Zhang},\ and\ \citenamefont
  {Li}}]{Zhang2014}%
  \BibitemOpen
  \bibfield  {author} {\bibinfo {author} {\bibfnamefont {W.}~\bibnamefont
  {Zhang}}, \bibinfo {author} {\bibfnamefont {Z.}~\bibnamefont {Huang}},
  \bibinfo {author} {\bibfnamefont {W.}~\bibnamefont {Zhang}},\ and\ \bibinfo
  {author} {\bibfnamefont {Y.}~\bibnamefont {Li}},\ }\href
  {https://doi.org/10.1007/s12274-014-0532-x} {\bibfield  {journal} {\bibinfo
  {journal} {Nano Research}\ }\textbf {\bibinfo {volume} {7}},\ \bibinfo
  {pages} {1731} (\bibinfo {year} {2014})}\BibitemShut {NoStop}%
\bibitem [{\citenamefont {Kanazawa}\ \emph {et~al.}(2016)\citenamefont
  {Kanazawa}, \citenamefont {Amemiya}, \citenamefont {Ishikawa}, \citenamefont
  {Upadhyaya}, \citenamefont {Tsuruta}, \citenamefont {Tanaka},\ and\
  \citenamefont {Miyamoto}}]{kanazawa2016}%
  \BibitemOpen
  \bibfield  {author} {\bibinfo {author} {\bibfnamefont {T.}~\bibnamefont
  {Kanazawa}}, \bibinfo {author} {\bibfnamefont {T.}~\bibnamefont {Amemiya}},
  \bibinfo {author} {\bibfnamefont {A.}~\bibnamefont {Ishikawa}}, \bibinfo
  {author} {\bibfnamefont {V.}~\bibnamefont {Upadhyaya}}, \bibinfo {author}
  {\bibfnamefont {K.}~\bibnamefont {Tsuruta}}, \bibinfo {author} {\bibfnamefont
  {T.}~\bibnamefont {Tanaka}},\ and\ \bibinfo {author} {\bibfnamefont
  {Y.}~\bibnamefont {Miyamoto}},\ }\href {https://doi.org/10.1038/srep22277}
  {\bibfield  {journal} {\bibinfo  {journal} {Scientific Reports}\ }\textbf
  {\bibinfo {volume} {6}},\ \bibinfo {pages} {22277} (\bibinfo {year}
  {2016})}\BibitemShut {NoStop}%
\bibitem [{\citenamefont {Singh}\ \emph {et~al.}(2022)\citenamefont {Singh},
  \citenamefont {Shao}, \citenamefont {Chen}, \citenamefont {Wu},\ and\
  \citenamefont {Tsai}}]{jitendra2022}%
  \BibitemOpen
  \bibfield  {author} {\bibinfo {author} {\bibfnamefont {J.}~\bibnamefont
  {Singh}}, \bibinfo {author} {\bibfnamefont {J.~H.}\ \bibnamefont {Shao}},
  \bibinfo {author} {\bibfnamefont {G.~T.}\ \bibnamefont {Chen}}, \bibinfo
  {author} {\bibfnamefont {H.~S.}\ \bibnamefont {Wu}},\ and\ \bibinfo {author}
  {\bibfnamefont {M.~L.}\ \bibnamefont {Tsai}},\ }\href
  {https://doi.org/10.1039/d2na00578f} {\bibfield  {journal} {\bibinfo
  {journal} {Nanoscale Advances}\ }\textbf {\bibinfo {volume} {5}},\ \bibinfo
  {pages} {171} (\bibinfo {year} {2022})}\BibitemShut {NoStop}%
\bibitem [{\citenamefont {Singh}\ and\ \citenamefont
  {Ahuja}(2019)}]{deobrat2019}%
  \BibitemOpen
  \bibfield  {author} {\bibinfo {author} {\bibfnamefont {D.}~\bibnamefont
  {Singh}}\ and\ \bibinfo {author} {\bibfnamefont {R.}~\bibnamefont {Ahuja}},\
  }\href {https://doi.org/10.1021/acsaem.9b01402} {\bibfield  {journal}
  {\bibinfo  {journal} {ACS Applied Energy Materials}\ }\textbf {\bibinfo
  {volume} {2}},\ \bibinfo {pages} {6891} (\bibinfo {year} {2019})}\BibitemShut
  {NoStop}%
\bibitem [{\citenamefont {Peng}\ \emph {et~al.}(2021)\citenamefont {Peng},
  \citenamefont {Najmaei}, \citenamefont {Dubey},\ and\ \citenamefont
  {Chung}}]{peng2021}%
  \BibitemOpen
  \bibfield  {author} {\bibinfo {author} {\bibfnamefont {J.}~\bibnamefont
  {Peng}}, \bibinfo {author} {\bibfnamefont {S.}~\bibnamefont {Najmaei}},
  \bibinfo {author} {\bibfnamefont {M.}~\bibnamefont {Dubey}},\ and\ \bibinfo
  {author} {\bibfnamefont {P.~W.}\ \bibnamefont {Chung}},\ }\href
  {https://doi.org/https://doi.org/10.1016/j.mtcomm.2020.101722} {\bibfield
  {journal} {\bibinfo  {journal} {Materials Today Communications}\ }\textbf
  {\bibinfo {volume} {26}},\ \bibinfo {pages} {101722} (\bibinfo {year}
  {2021})}\BibitemShut {NoStop}%
\bibitem [{\citenamefont {Ib{\'a}{\~n}ez}\ \emph {et~al.}(2018)\citenamefont
  {Ib{\'a}{\~n}ez}, \citenamefont {Wo{\'z}niak}, \citenamefont {Dybala},
  \citenamefont {Oliva}, \citenamefont {Hern{\'a}ndez},\ and\ \citenamefont
  {Kudrawiec}}]{ibanez2018high}%
  \BibitemOpen
  \bibfield  {author} {\bibinfo {author} {\bibfnamefont {J.}~\bibnamefont
  {Ib{\'a}{\~n}ez}}, \bibinfo {author} {\bibfnamefont {T.}~\bibnamefont
  {Wo{\'z}niak}}, \bibinfo {author} {\bibfnamefont {F.}~\bibnamefont {Dybala}},
  \bibinfo {author} {\bibfnamefont {R.}~\bibnamefont {Oliva}}, \bibinfo
  {author} {\bibfnamefont {S.}~\bibnamefont {Hern{\'a}ndez}},\ and\ \bibinfo
  {author} {\bibfnamefont {R.}~\bibnamefont {Kudrawiec}},\ }\href
  {https://doi.org/10.1038/s41598-018-31051-y} {\bibfield  {journal} {\bibinfo
  {journal} {Scientific Reports}\ }\textbf {\bibinfo {volume} {8}},\ \bibinfo
  {pages} {1} (\bibinfo {year} {2018})}\BibitemShut {NoStop}%
\bibitem [{\citenamefont {Grzeszczyk}\ \emph {et~al.}(2022)\citenamefont
  {Grzeszczyk}, \citenamefont {Gawraczyński}, \citenamefont {Woźniak},
  \citenamefont {Ibáñez}, \citenamefont {Muhammad}, \citenamefont {Zhao},
  \citenamefont {Molas},\ and\ \citenamefont {Babiński}}]{Grzeszczyk2022}%
  \BibitemOpen
  \bibfield  {author} {\bibinfo {author} {\bibfnamefont {M.}~\bibnamefont
  {Grzeszczyk}}, \bibinfo {author} {\bibfnamefont {J.}~\bibnamefont
  {Gawraczyński}}, \bibinfo {author} {\bibfnamefont {T.}~\bibnamefont
  {Woźniak}}, \bibinfo {author} {\bibfnamefont {J.}~\bibnamefont {Ibáñez}},
  \bibinfo {author} {\bibfnamefont {Z.}~\bibnamefont {Muhammad}}, \bibinfo
  {author} {\bibfnamefont {W.}~\bibnamefont {Zhao}}, \bibinfo {author}
  {\bibfnamefont {M.}~\bibnamefont {Molas}},\ and\ \bibinfo {author}
  {\bibfnamefont {A.}~\bibnamefont {Babiński}},\ }\href
  {https://doi.org/10.12693/APhysPolA.141.95} {\bibfield  {journal} {\bibinfo
  {journal} {Acta Physica Polonica A}\ }\textbf {\bibinfo {volume} {141}},\
  \bibinfo {pages} {95} (\bibinfo {year} {2022})}\BibitemShut {NoStop}%
\bibitem [{\citenamefont {Zawadzka}\ \emph {et~al.}(2023)\citenamefont
  {Zawadzka}, \citenamefont {Wo{\'{z}}niak}, \citenamefont {Strawski},
  \citenamefont {Antoniazzi}, \citenamefont {Grzeszczyk}, \citenamefont
  {Olkowska-Pucko}, \citenamefont {Muhammad}, \citenamefont {Ibanez},
  \citenamefont {Zhao}, \citenamefont {Jadczak}, \citenamefont
  {St{\c{e}}pniewski}, \citenamefont {Babi{\'{n}}ski},\ and\ \citenamefont
  {Molas}}]{Zawadzka2023}%
  \BibitemOpen
  \bibfield  {author} {\bibinfo {author} {\bibfnamefont {N.}~\bibnamefont
  {Zawadzka}}, \bibinfo {author} {\bibfnamefont {T.}~\bibnamefont
  {Wo{\'{z}}niak}}, \bibinfo {author} {\bibfnamefont {M.}~\bibnamefont
  {Strawski}}, \bibinfo {author} {\bibfnamefont {I.}~\bibnamefont
  {Antoniazzi}}, \bibinfo {author} {\bibfnamefont {M.}~\bibnamefont
  {Grzeszczyk}}, \bibinfo {author} {\bibfnamefont {K.}~\bibnamefont
  {Olkowska-Pucko}}, \bibinfo {author} {\bibfnamefont {Z.}~\bibnamefont
  {Muhammad}}, \bibinfo {author} {\bibfnamefont {J.}~\bibnamefont {Ibanez}},
  \bibinfo {author} {\bibfnamefont {W.}~\bibnamefont {Zhao}}, \bibinfo {author}
  {\bibfnamefont {J.}~\bibnamefont {Jadczak}}, \bibinfo {author} {\bibfnamefont
  {R.}~\bibnamefont {St{\c{e}}pniewski}}, \bibinfo {author} {\bibfnamefont
  {A.}~\bibnamefont {Babi{\'{n}}ski}},\ and\ \bibinfo {author} {\bibfnamefont
  {M.~R.}\ \bibnamefont {Molas}},\ }\href {https://doi.org/10.1063/5.0126894}
  {\bibfield  {journal} {\bibinfo  {journal} {Applied Physics Letters}\
  }\textbf {\bibinfo {volume} {122}},\ \bibinfo {pages} {042102} (\bibinfo
  {year} {2023})}\BibitemShut {NoStop}%
\bibitem [{\citenamefont {Antoniazzi}\ \emph {et~al.}(2023)\citenamefont
  {Antoniazzi}, \citenamefont {Zawadzka}, \citenamefont {Grzeszczyk},
  \citenamefont {Wo{\'{z}}niak}, \citenamefont {Ib{\'{a}}{\~{n}}ez},
  \citenamefont {Muhammad}, \citenamefont {Zhao}, \citenamefont {Molas},\ and\
  \citenamefont {Babi{\'{n}}ski}}]{antoniazzi2023}%
  \BibitemOpen
  \bibfield  {author} {\bibinfo {author} {\bibfnamefont {I.}~\bibnamefont
  {Antoniazzi}}, \bibinfo {author} {\bibfnamefont {N.}~\bibnamefont
  {Zawadzka}}, \bibinfo {author} {\bibfnamefont {M.}~\bibnamefont
  {Grzeszczyk}}, \bibinfo {author} {\bibfnamefont {T.}~\bibnamefont
  {Wo{\'{z}}niak}}, \bibinfo {author} {\bibfnamefont {J.}~\bibnamefont
  {Ib{\'{a}}{\~{n}}ez}}, \bibinfo {author} {\bibfnamefont {Z.}~\bibnamefont
  {Muhammad}}, \bibinfo {author} {\bibfnamefont {W.}~\bibnamefont {Zhao}},
  \bibinfo {author} {\bibfnamefont {M.~R.}\ \bibnamefont {Molas}},\ and\
  \bibinfo {author} {\bibfnamefont {A.}~\bibnamefont {Babi{\'{n}}ski}},\ }\href
  {https://iopscience.iop.org/article/10.1088/1361-648X/acce18} {\bibfield
  {journal} {\bibinfo  {journal} {Journal of Physics: Condensed MatterMatter}\
  }\textbf {\bibinfo {volume} {35}},\ \bibinfo {pages} {3055401} (\bibinfo
  {year} {2023})}\BibitemShut {NoStop}%
\bibitem [{\citenamefont {Zhong}\ \emph {et~al.}(2023)\citenamefont {Zhong},
  \citenamefont {Deng}, \citenamefont {Hong},\ and\ \citenamefont
  {Yue}}]{zhong2023}%
  \BibitemOpen
  \bibfield  {author} {\bibinfo {author} {\bibfnamefont {W.}~\bibnamefont
  {Zhong}}, \bibinfo {author} {\bibfnamefont {W.}~\bibnamefont {Deng}},
  \bibinfo {author} {\bibfnamefont {F.}~\bibnamefont {Hong}},\ and\ \bibinfo
  {author} {\bibfnamefont {B.}~\bibnamefont {Yue}},\ }\href
  {https://doi.org/10.1103/PhysRevB.107.134118} {\bibfield  {journal} {\bibinfo
   {journal} {Physical Review B}\ }\textbf {\bibinfo {volume} {107}},\ \bibinfo
  {pages} {1} (\bibinfo {year} {2023})}\BibitemShut {NoStop}%
\bibitem [{\citenamefont {Hong}\ \emph {et~al.}(2022)\citenamefont {Hong},
  \citenamefont {Dai}, \citenamefont {Hu}, \citenamefont {Zhang}, \citenamefont
  {Li},\ and\ \citenamefont {He}}]{hong2022}%
  \BibitemOpen
  \bibfield  {author} {\bibinfo {author} {\bibfnamefont {M.}~\bibnamefont
  {Hong}}, \bibinfo {author} {\bibfnamefont {L.}~\bibnamefont {Dai}}, \bibinfo
  {author} {\bibfnamefont {H.}~\bibnamefont {Hu}}, \bibinfo {author}
  {\bibfnamefont {X.}~\bibnamefont {Zhang}}, \bibinfo {author} {\bibfnamefont
  {C.}~\bibnamefont {Li}},\ and\ \bibinfo {author} {\bibfnamefont
  {Y.}~\bibnamefont {He}},\ }\href {https://doi.org/10.1039/D2TC01669A}
  {\bibfield  {journal} {\bibinfo  {journal} {Journal of Materials Chemistry
  C}\ }\textbf {\bibinfo {volume} {10}},\ \bibinfo {pages} {10541} (\bibinfo
  {year} {2022})}\BibitemShut {NoStop}%
\bibitem [{\citenamefont {Zhang}\ \emph {et~al.}(2023)\citenamefont {Zhang},
  \citenamefont {Wang}, \citenamefont {Liu}, \citenamefont {Zhen},
  \citenamefont {Wan}, \citenamefont {Deng}, \citenamefont {Han}, \citenamefont
  {Chen},\ and\ \citenamefont {Gao}}]{Zhang2023}%
  \BibitemOpen
  \bibfield  {author} {\bibinfo {author} {\bibfnamefont {S.}~\bibnamefont
  {Zhang}}, \bibinfo {author} {\bibfnamefont {H.}~\bibnamefont {Wang}},
  \bibinfo {author} {\bibfnamefont {H.}~\bibnamefont {Liu}}, \bibinfo {author}
  {\bibfnamefont {J.}~\bibnamefont {Zhen}}, \bibinfo {author} {\bibfnamefont
  {S.}~\bibnamefont {Wan}}, \bibinfo {author} {\bibfnamefont {W.}~\bibnamefont
  {Deng}}, \bibinfo {author} {\bibfnamefont {Y.}~\bibnamefont {Han}}, \bibinfo
  {author} {\bibfnamefont {B.}~\bibnamefont {Chen}},\ and\ \bibinfo {author}
  {\bibfnamefont {C.}~\bibnamefont {Gao}},\ }\href
  {https://doi.org/10.1103/PhysRevMaterials.7.104802} {\bibfield  {journal}
  {\bibinfo  {journal} {Phys. Rev. Mater.}\ }\textbf {\bibinfo {volume} {7}},\
  \bibinfo {pages} {104802} (\bibinfo {year} {2023})}\BibitemShut {NoStop}%
\bibitem [{\citenamefont {Zhang}\ \emph {et~al.}(2022)\citenamefont {Zhang},
  \citenamefont {Fang}, \citenamefont {Wu}, \citenamefont {Cao}, \citenamefont
  {Zhou}, \citenamefont {Zhu},\ and\ \citenamefont {Wu}}]{Zhang_2022}%
  \BibitemOpen
  \bibfield  {author} {\bibinfo {author} {\bibfnamefont {R.}~\bibnamefont
  {Zhang}}, \bibinfo {author} {\bibfnamefont {Y.}~\bibnamefont {Fang}},
  \bibinfo {author} {\bibfnamefont {F.}~\bibnamefont {Wu}}, \bibinfo {author}
  {\bibfnamefont {X.}~\bibnamefont {Cao}}, \bibinfo {author} {\bibfnamefont
  {Y.}~\bibnamefont {Zhou}}, \bibinfo {author} {\bibfnamefont {Z.-Z.}\
  \bibnamefont {Zhu}},\ and\ \bibinfo {author} {\bibfnamefont {S.}~\bibnamefont
  {Wu}},\ }\href {https://doi.org/10.1088/1361-6463/ac6401} {\bibfield
  {journal} {\bibinfo  {journal} {Journal of Physics D: Applied Physics}\
  }\textbf {\bibinfo {volume} {55}},\ \bibinfo {pages} {295304} (\bibinfo
  {year} {2022})}\BibitemShut {NoStop}%
\bibitem [{\citenamefont {Lucovsky}\ \emph {et~al.}(1973)\citenamefont
  {Lucovsky}, \citenamefont {White}, \citenamefont {Benda},\ and\ \citenamefont
  {Revelli}}]{lukovsky1973IR}%
  \BibitemOpen
  \bibfield  {author} {\bibinfo {author} {\bibfnamefont {G.}~\bibnamefont
  {Lucovsky}}, \bibinfo {author} {\bibfnamefont {R.~M.}\ \bibnamefont {White}},
  \bibinfo {author} {\bibfnamefont {J.~A.}\ \bibnamefont {Benda}},\ and\
  \bibinfo {author} {\bibfnamefont {J.~F.}\ \bibnamefont {Revelli}},\ }\href
  {https://doi.org/10.1103/PhysRevB.7.3859} {\bibfield  {journal} {\bibinfo
  {journal} {Phys. Rev. B}\ }\textbf {\bibinfo {volume} {7}},\ \bibinfo {pages}
  {3859} (\bibinfo {year} {1973})}\BibitemShut {NoStop}%
\bibitem [{\citenamefont {Iwasaki}\ \emph {et~al.}(1982)\citenamefont
  {Iwasaki}, \citenamefont {Kuroda},\ and\ \citenamefont
  {Y}}]{iwasaki1982Raman}%
  \BibitemOpen
  \bibfield  {author} {\bibinfo {author} {\bibfnamefont {T.}~\bibnamefont
  {Iwasaki}}, \bibinfo {author} {\bibfnamefont {N.}~\bibnamefont {Kuroda}},\
  and\ \bibinfo {author} {\bibfnamefont {N.}~\bibnamefont {Y}},\ }\href
  {https://doi.org/10.1143/JPSJ.51.2233} {\bibfield  {journal} {\bibinfo
  {journal} {J. Phys. Soc. Jpn.}\ }\textbf {\bibinfo {volume} {52}},\ \bibinfo
  {pages} {2233} (\bibinfo {year} {1982})}\BibitemShut {NoStop}%
\bibitem [{\citenamefont {Cingolani}\ \emph {et~al.}(1987)\citenamefont
  {Cingolani}, \citenamefont {Lugara}, \citenamefont {Scamarcio},\ and\
  \citenamefont {L{\'e}vy}}]{cingolani1987raman}%
  \BibitemOpen
  \bibfield  {author} {\bibinfo {author} {\bibfnamefont {A.}~\bibnamefont
  {Cingolani}}, \bibinfo {author} {\bibfnamefont {M.}~\bibnamefont {Lugara}},
  \bibinfo {author} {\bibfnamefont {G.}~\bibnamefont {Scamarcio}},\ and\
  \bibinfo {author} {\bibfnamefont {F.}~\bibnamefont {L{\'e}vy}},\ }\href
  {https://doi.org/10.1016/0038-1098(87)91126-4} {\bibfield  {journal}
  {\bibinfo  {journal} {Solid State Communications}\ }\textbf {\bibinfo
  {volume} {62}},\ \bibinfo {pages} {121} (\bibinfo {year} {1987})}\BibitemShut
  {NoStop}%
\bibitem [{\citenamefont {Roubi}\ and\ \citenamefont
  {Carlone}(1988)}]{roubi1988resonance}%
  \BibitemOpen
  \bibfield  {author} {\bibinfo {author} {\bibfnamefont {L.}~\bibnamefont
  {Roubi}}\ and\ \bibinfo {author} {\bibfnamefont {C.}~\bibnamefont
  {Carlone}},\ }\href {https://doi.org/10.1103/PhysRevB.37.6808} {\bibfield
  {journal} {\bibinfo  {journal} {Physical Review B}\ }\textbf {\bibinfo
  {volume} {37}},\ \bibinfo {pages} {6808} (\bibinfo {year}
  {1988})}\BibitemShut {NoStop}%
\bibitem [{\citenamefont {Klotz}\ \emph {et~al.}(2009)\citenamefont {Klotz},
  \citenamefont {Chervin}, \citenamefont {Munsch},\ and\ \citenamefont
  {Marchand}}]{Klotz_2009}%
  \BibitemOpen
  \bibfield  {author} {\bibinfo {author} {\bibfnamefont {S.}~\bibnamefont
  {Klotz}}, \bibinfo {author} {\bibfnamefont {J.-C.}\ \bibnamefont {Chervin}},
  \bibinfo {author} {\bibfnamefont {P.}~\bibnamefont {Munsch}},\ and\ \bibinfo
  {author} {\bibfnamefont {G.~L.}\ \bibnamefont {Marchand}},\ }\href
  {https://doi.org/10.1088/0022-3727/42/7/075413} {\bibfield  {journal}
  {\bibinfo  {journal} {Journal of Physics D: Applied Physics}\ }\textbf
  {\bibinfo {volume} {42}},\ \bibinfo {pages} {075413} (\bibinfo {year}
  {2009})}\BibitemShut {NoStop}%
\bibitem [{\citenamefont {Porezag}\ and\ \citenamefont
  {Pederson}(1996)}]{Porezag1996}%
  \BibitemOpen
  \bibfield  {author} {\bibinfo {author} {\bibfnamefont {D.}~\bibnamefont
  {Porezag}}\ and\ \bibinfo {author} {\bibfnamefont {M.~R.}\ \bibnamefont
  {Pederson}},\ }\href {https://doi.org/10.1103/PhysRevB.54.7830} {\bibfield
  {journal} {\bibinfo  {journal} {Physical Review B - Condensed Matter and
  Materials Physics}\ }\textbf {\bibinfo {volume} {54}},\ \bibinfo {pages}
  {7830} (\bibinfo {year} {1996})}\BibitemShut {NoStop}%
\bibitem [{\citenamefont {Carvalho}\ \emph {et~al.}(2015)\citenamefont
  {Carvalho}, \citenamefont {Malard}, \citenamefont {Alves}, \citenamefont
  {Fantini},\ and\ \citenamefont {Pimenta}}]{Carvalho2015}%
  \BibitemOpen
  \bibfield  {author} {\bibinfo {author} {\bibfnamefont {B.~R.}\ \bibnamefont
  {Carvalho}}, \bibinfo {author} {\bibfnamefont {L.~M.}\ \bibnamefont
  {Malard}}, \bibinfo {author} {\bibfnamefont {J.~M.}\ \bibnamefont {Alves}},
  \bibinfo {author} {\bibfnamefont {C.}~\bibnamefont {Fantini}},\ and\ \bibinfo
  {author} {\bibfnamefont {M.~A.}\ \bibnamefont {Pimenta}},\ }\href
  {https://doi.org/10.1103/PhysRevLett.114.136403} {\bibfield  {journal}
  {\bibinfo  {journal} {Phys. Rev. Lett.}\ }\textbf {\bibinfo {volume} {114}},\
  \bibinfo {pages} {136403} (\bibinfo {year} {2015})}\BibitemShut {NoStop}%
\bibitem [{\citenamefont {Steele}\ \emph {et~al.}(2016)\citenamefont {Steele},
  \citenamefont {Puech},\ and\ \citenamefont {Lewis}}]{steele2016}%
  \BibitemOpen
  \bibfield  {author} {\bibinfo {author} {\bibfnamefont {J.~A.}\ \bibnamefont
  {Steele}}, \bibinfo {author} {\bibfnamefont {P.}~\bibnamefont {Puech}},\ and\
  \bibinfo {author} {\bibfnamefont {R.~A.}\ \bibnamefont {Lewis}},\ }\bibfield
  {journal} {\bibinfo  {journal} {Journal of Applied Physics}\ }\textbf
  {\bibinfo {volume} {120}},\ \href {https://doi.org/10.1063/1.4959824}
  {10.1063/1.4959824} (\bibinfo {year} {2016})\BibitemShut {NoStop}%
\bibitem [{\citenamefont {Cardona}\ and\ \citenamefont
  {Guntherodt}(1982)}]{Cardona1982}%
  \BibitemOpen
  \bibfield  {author} {\bibinfo {author} {\bibfnamefont {M.}~\bibnamefont
  {Cardona}}\ and\ \bibinfo {author} {\bibfnamefont {G.}~\bibnamefont
  {Guntherodt}},\ }\href@noop {} {\bibfield  {journal} {\bibinfo  {journal}
  {Topics in Applied Physics}\ }\textbf {\bibinfo {volume} {50}} (\bibinfo
  {year} {1982})}\BibitemShut {NoStop}%
\bibitem [{\citenamefont {Pimenta}\ \emph {et~al.}(2021)\citenamefont
  {Pimenta}, \citenamefont {Resende}, \citenamefont {Ribeiro},\ and\
  \citenamefont {Carvalho}}]{Pimenta2021}%
  \BibitemOpen
  \bibfield  {author} {\bibinfo {author} {\bibfnamefont {M.~A.}\ \bibnamefont
  {Pimenta}}, \bibinfo {author} {\bibfnamefont {G.~C.}\ \bibnamefont
  {Resende}}, \bibinfo {author} {\bibfnamefont {H.~B.}\ \bibnamefont
  {Ribeiro}},\ and\ \bibinfo {author} {\bibfnamefont {B.~R.}\ \bibnamefont
  {Carvalho}},\ }\href {https://doi.org/10.1039/D1CP03626B} {\bibfield
  {journal} {\bibinfo  {journal} {Phys. Chem. Chem. Phys.}\ }\textbf {\bibinfo
  {volume} {23}},\ \bibinfo {pages} {27103} (\bibinfo {year}
  {2021})}\BibitemShut {NoStop}%
\bibitem [{\citenamefont {Kranert}\ \emph {et~al.}(2016)\citenamefont
  {Kranert}, \citenamefont {Sturm}, \citenamefont {Schmidt-Grund},\ and\
  \citenamefont {Grundmann}}]{kranert2016}%
  \BibitemOpen
  \bibfield  {author} {\bibinfo {author} {\bibfnamefont {C.}~\bibnamefont
  {Kranert}}, \bibinfo {author} {\bibfnamefont {C.}~\bibnamefont {Sturm}},
  \bibinfo {author} {\bibfnamefont {R.}~\bibnamefont {Schmidt-Grund}},\ and\
  \bibinfo {author} {\bibfnamefont {M.}~\bibnamefont {Grundmann}},\ }\bibfield
  {journal} {\bibinfo  {journal} {Physical Review Letters}\ }\textbf {\bibinfo
  {volume} {116}},\ \href {https://doi.org/10.1103/PhysRevLett.116.127401}
  {10.1103/PhysRevLett.116.127401} (\bibinfo {year} {2016})\BibitemShut
  {NoStop}%
\bibitem [{\citenamefont {Song}\ \emph {et~al.}(2016)\citenamefont {Song},
  \citenamefont {Pan}, \citenamefont {Wang},\ and\ \citenamefont
  {et~al.}}]{Song2016}%
  \BibitemOpen
  \bibfield  {author} {\bibinfo {author} {\bibfnamefont {Q.}~\bibnamefont
  {Song}}, \bibinfo {author} {\bibfnamefont {X.}~\bibnamefont {Pan}}, \bibinfo
  {author} {\bibfnamefont {H.}~\bibnamefont {Wang}},\ and\ \bibinfo {author}
  {\bibnamefont {et~al.}},\ }\href {https://doi.org/10.1038/srep29254}
  {\bibfield  {journal} {\bibinfo  {journal} {Scientific Reports}\ }\textbf
  {\bibinfo {volume} {6}},\ \bibinfo {pages} {29254} (\bibinfo {year}
  {2016})}\BibitemShut {NoStop}%
\bibitem [{\citenamefont {Ribeiro}\ \emph {et~al.}(2015)\citenamefont
  {Ribeiro}, \citenamefont {Pimenta}, \citenamefont {Matos}, \citenamefont
  {Moreira}, \citenamefont {Rodin}, \citenamefont {Zapata}, \citenamefont
  {Souza},\ and\ \citenamefont {Neto}}]{Ribeiro2015}%
  \BibitemOpen
  \bibfield  {author} {\bibinfo {author} {\bibfnamefont {H.~B.}\ \bibnamefont
  {Ribeiro}}, \bibinfo {author} {\bibfnamefont {M.~A.}\ \bibnamefont
  {Pimenta}}, \bibinfo {author} {\bibfnamefont {C.~J.~D.}\ \bibnamefont
  {Matos}}, \bibinfo {author} {\bibfnamefont {R.~L.}\ \bibnamefont {Moreira}},
  \bibinfo {author} {\bibfnamefont {A.~S.}\ \bibnamefont {Rodin}}, \bibinfo
  {author} {\bibfnamefont {J.~D.}\ \bibnamefont {Zapata}}, \bibinfo {author}
  {\bibfnamefont {E.~A.~D.}\ \bibnamefont {Souza}},\ and\ \bibinfo {author}
  {\bibfnamefont {A.~H.~C.}\ \bibnamefont {Neto}},\ }\href
  {https://doi.org/10.1021/acsnano.5b00698} {\bibfield  {journal} {\bibinfo
  {journal} {ACS Nano}\ }\textbf {\bibinfo {volume} {9}},\ \bibinfo {pages}
  {4270} (\bibinfo {year} {2015})}\BibitemShut {NoStop}%
\bibitem [{\citenamefont {Wang}\ \emph {et~al.}(2021)\citenamefont {Wang},
  \citenamefont {Chen}, \citenamefont {Guo}, \citenamefont {Liu}, \citenamefont
  {Jiang}, \citenamefont {Zheng}, \citenamefont {Wang}, \citenamefont
  {Al-Makeen}, \citenamefont {Ouyang}, \citenamefont {Xia},\ and\ \citenamefont
  {Huang}}]{Wang2021}%
  \BibitemOpen
  \bibfield  {author} {\bibinfo {author} {\bibfnamefont {Y.}~\bibnamefont
  {Wang}}, \bibinfo {author} {\bibfnamefont {F.}~\bibnamefont {Chen}}, \bibinfo
  {author} {\bibfnamefont {X.}~\bibnamefont {Guo}}, \bibinfo {author}
  {\bibfnamefont {J.}~\bibnamefont {Liu}}, \bibinfo {author} {\bibfnamefont
  {J.}~\bibnamefont {Jiang}}, \bibinfo {author} {\bibfnamefont
  {X.}~\bibnamefont {Zheng}}, \bibinfo {author} {\bibfnamefont
  {Z.}~\bibnamefont {Wang}}, \bibinfo {author} {\bibfnamefont {M.~M.}\
  \bibnamefont {Al-Makeen}}, \bibinfo {author} {\bibfnamefont {F.}~\bibnamefont
  {Ouyang}}, \bibinfo {author} {\bibfnamefont {Q.}~\bibnamefont {Xia}},\ and\
  \bibinfo {author} {\bibfnamefont {H.}~\bibnamefont {Huang}},\ }\href
  {https://doi.org/10.1021/acs.jpclett.1c03218} {\bibfield  {journal} {\bibinfo
   {journal} {The Journal of Physical Chemistry Letters}\ }\textbf {\bibinfo
  {volume} {12}},\ \bibinfo {pages} {10753} (\bibinfo {year}
  {2021})}\BibitemShut {NoStop}%
\bibitem [{\citenamefont {Gong}\ \emph {et~al.}(2022)\citenamefont {Gong},
  \citenamefont {Zhao}, \citenamefont {Zhou}, \citenamefont {Li}, \citenamefont
  {Mao}, \citenamefont {Bao}, \citenamefont {Zhang},\ and\ \citenamefont
  {Wang}}]{Gong2022}%
  \BibitemOpen
  \bibfield  {author} {\bibinfo {author} {\bibfnamefont {Y.}~\bibnamefont
  {Gong}}, \bibinfo {author} {\bibfnamefont {Y.}~\bibnamefont {Zhao}}, \bibinfo
  {author} {\bibfnamefont {Z.}~\bibnamefont {Zhou}}, \bibinfo {author}
  {\bibfnamefont {D.}~\bibnamefont {Li}}, \bibinfo {author} {\bibfnamefont
  {H.}~\bibnamefont {Mao}}, \bibinfo {author} {\bibfnamefont {Q.}~\bibnamefont
  {Bao}}, \bibinfo {author} {\bibfnamefont {Y.}~\bibnamefont {Zhang}},\ and\
  \bibinfo {author} {\bibfnamefont {G.~P.}\ \bibnamefont {Wang}},\ }\href
  {https://doi.org/10.1002/adom.202200038} {\bibfield  {journal} {\bibinfo
  {journal} {Advanced Optical Materials}\ }\textbf {\bibinfo {volume} {10}},\
  \bibinfo {pages} {2200038} (\bibinfo {year} {2022})}\BibitemShut {NoStop}%
\bibitem [{\citenamefont {Chenet}\ \emph {et~al.}(2015)\citenamefont {Chenet},
  \citenamefont {Aslan}, \citenamefont {Huang}, \citenamefont {Fan},
  \citenamefont {van~der Zande}, \citenamefont {Heinz},\ and\ \citenamefont
  {Hone}}]{Chenet2015}%
  \BibitemOpen
  \bibfield  {author} {\bibinfo {author} {\bibfnamefont {D.~A.}\ \bibnamefont
  {Chenet}}, \bibinfo {author} {\bibfnamefont {B.}~\bibnamefont {Aslan}},
  \bibinfo {author} {\bibfnamefont {P.~Y.}\ \bibnamefont {Huang}}, \bibinfo
  {author} {\bibfnamefont {C.}~\bibnamefont {Fan}}, \bibinfo {author}
  {\bibfnamefont {A.~M.}\ \bibnamefont {van~der Zande}}, \bibinfo {author}
  {\bibfnamefont {T.~F.}\ \bibnamefont {Heinz}},\ and\ \bibinfo {author}
  {\bibfnamefont {J.~C.}\ \bibnamefont {Hone}},\ }\href
  {https://doi.org/10.1021/acs.nanolett.5b00910} {\bibfield  {journal}
  {\bibinfo  {journal} {Nano Letters}\ }\textbf {\bibinfo {volume} {15}},\
  \bibinfo {pages} {5667} (\bibinfo {year} {2015})}\BibitemShut {NoStop}%
\bibitem [{\citenamefont {Zawadzka}\ \emph {et~al.}(2021)\citenamefont
  {Zawadzka}, \citenamefont {Kipczak}, \citenamefont {Wo\'zniak}, \citenamefont
  {Olkowska-Pucko}, \citenamefont {Grzeszczyk}, \citenamefont {Babi\'nski},\
  and\ \citenamefont {Molas}}]{Zawadzka2021}%
  \BibitemOpen
  \bibfield  {author} {\bibinfo {author} {\bibfnamefont {N.}~\bibnamefont
  {Zawadzka}}, \bibinfo {author} {\bibfnamefont {L.}~\bibnamefont {Kipczak}},
  \bibinfo {author} {\bibfnamefont {T.}~\bibnamefont {Wo\'zniak}}, \bibinfo
  {author} {\bibfnamefont {K.}~\bibnamefont {Olkowska-Pucko}}, \bibinfo
  {author} {\bibfnamefont {M.}~\bibnamefont {Grzeszczyk}}, \bibinfo {author}
  {\bibfnamefont {A.}~\bibnamefont {Babi\'nski}},\ and\ \bibinfo {author}
  {\bibfnamefont {M.~R.}\ \bibnamefont {Molas}},\ }\href
  {https://doi.org/10.3390/nano11113109} {\bibfield  {journal} {\bibinfo
  {journal} {Nanomaterials}\ }\textbf {\bibinfo {volume} {11}},\ \bibinfo
  {pages} {3109} (\bibinfo {year} {2021})}\BibitemShut {NoStop}%
\bibitem [{\citenamefont {Lorchat}\ \emph {et~al.}(2016)\citenamefont
  {Lorchat}, \citenamefont {Froehlicher},\ and\ \citenamefont
  {Berciaud}}]{Lorchat2016}%
  \BibitemOpen
  \bibfield  {author} {\bibinfo {author} {\bibfnamefont {E.}~\bibnamefont
  {Lorchat}}, \bibinfo {author} {\bibfnamefont {G.}~\bibnamefont
  {Froehlicher}},\ and\ \bibinfo {author} {\bibfnamefont {S.}~\bibnamefont
  {Berciaud}},\ }\href {https://doi.org/10.1021/acsnano.5b07844} {\bibfield
  {journal} {\bibinfo  {journal} {ACS Nano}\ }\textbf {\bibinfo {volume}
  {10}},\ \bibinfo {pages} {2752} (\bibinfo {year} {2016})}\BibitemShut
  {NoStop}%
\bibitem [{\citenamefont {Froehlicher}\ \emph {et~al.}(2015)\citenamefont
  {Froehlicher}, \citenamefont {Lorchat}, \citenamefont {Fernique},
  \citenamefont {Joshi}, \citenamefont {Molina-Sánchez}, \citenamefont
  {Wirtz},\ and\ \citenamefont {Berciaud}}]{Froehlicher}%
  \BibitemOpen
  \bibfield  {author} {\bibinfo {author} {\bibfnamefont {G.}~\bibnamefont
  {Froehlicher}}, \bibinfo {author} {\bibfnamefont {E.}~\bibnamefont
  {Lorchat}}, \bibinfo {author} {\bibfnamefont {F.}~\bibnamefont {Fernique}},
  \bibinfo {author} {\bibfnamefont {C.}~\bibnamefont {Joshi}}, \bibinfo
  {author} {\bibfnamefont {A.}~\bibnamefont {Molina-Sánchez}}, \bibinfo
  {author} {\bibfnamefont {L.}~\bibnamefont {Wirtz}},\ and\ \bibinfo {author}
  {\bibfnamefont {S.}~\bibnamefont {Berciaud}},\ }\href
  {https://doi.org/10.1021/acs.nanolett.5b02683} {\bibfield  {journal}
  {\bibinfo  {journal} {Nano Letters}\ }\textbf {\bibinfo {volume} {15}},\
  \bibinfo {pages} {6481} (\bibinfo {year} {2015})}\BibitemShut {NoStop}%
\bibitem [{\citenamefont {Grzeszczyk}\ \emph {et~al.}(2016)\citenamefont
  {Grzeszczyk}, \citenamefont {Gołasa}, \citenamefont {Zinkiewicz},
  \citenamefont {Nogajewski}, \citenamefont {Molas}, \citenamefont {Potemski},
  \citenamefont {Wysmołek},\ and\ \citenamefont
  {Babiński}}]{grzeszczyk2016Raman}%
  \BibitemOpen
  \bibfield  {author} {\bibinfo {author} {\bibfnamefont {M.}~\bibnamefont
  {Grzeszczyk}}, \bibinfo {author} {\bibfnamefont {K.}~\bibnamefont {Gołasa}},
  \bibinfo {author} {\bibfnamefont {M.}~\bibnamefont {Zinkiewicz}}, \bibinfo
  {author} {\bibfnamefont {K.}~\bibnamefont {Nogajewski}}, \bibinfo {author}
  {\bibfnamefont {M.~R.}\ \bibnamefont {Molas}}, \bibinfo {author}
  {\bibfnamefont {M.}~\bibnamefont {Potemski}}, \bibinfo {author}
  {\bibfnamefont {A.}~\bibnamefont {Wysmołek}},\ and\ \bibinfo {author}
  {\bibfnamefont {A.}~\bibnamefont {Babiński}},\ }\href
  {https://doi.org/10.1088/2053-1583/3/2/025010} {\bibfield  {journal}
  {\bibinfo  {journal} {2D Materials}\ }\textbf {\bibinfo {volume} {3}},\
  \bibinfo {pages} {025010} (\bibinfo {year} {2016})}\BibitemShut {NoStop}%
\bibitem [{\citenamefont {Łacińska}\ \emph {et~al.}(2022)\citenamefont
  {Łacińska}, \citenamefont {Furman}, \citenamefont {Binder}, \citenamefont
  {Lutsyk}, \citenamefont {Kowalczyk}, \citenamefont {Stepniewski},\ and\
  \citenamefont {Wysmolek}}]{Lacinska2022}%
  \BibitemOpen
  \bibfield  {author} {\bibinfo {author} {\bibfnamefont {E.~M.}\ \bibnamefont
  {Łacińska}}, \bibinfo {author} {\bibfnamefont {M.}~\bibnamefont {Furman}},
  \bibinfo {author} {\bibfnamefont {J.}~\bibnamefont {Binder}}, \bibinfo
  {author} {\bibfnamefont {I.}~\bibnamefont {Lutsyk}}, \bibinfo {author}
  {\bibfnamefont {P.~J.}\ \bibnamefont {Kowalczyk}}, \bibinfo {author}
  {\bibfnamefont {R.}~\bibnamefont {Stepniewski}},\ and\ \bibinfo {author}
  {\bibfnamefont {A.}~\bibnamefont {Wysmolek}},\ }\href
  {https://doi.org/10.1021/acs.nanolett.1c04990} {\bibfield  {journal}
  {\bibinfo  {journal} {Nano Letters}\ }\textbf {\bibinfo {volume} {22}},\
  \bibinfo {pages} {2835} (\bibinfo {year} {2022})}\BibitemShut {NoStop}%
\bibitem [{\citenamefont {Koningstein}(1969)}]{Koninstein1969}%
  \BibitemOpen
  \bibfield  {author} {\bibinfo {author} {\bibfnamefont {J.~A.}\ \bibnamefont
  {Koningstein}},\ }\href {https://doi.org/10.1063/1.1672118} {\bibfield
  {journal} {\bibinfo  {journal} {The Journal of Chemical Physics}\ }\textbf
  {\bibinfo {volume} {51}},\ \bibinfo {pages} {1163} (\bibinfo {year}
  {1969})}\BibitemShut {NoStop}%
\bibitem [{\citenamefont {Ib{\'a}{\~n}ez}()}]{ibanezPrivate}%
  \BibitemOpen
  \bibfield  {author} {\bibinfo {author} {\bibfnamefont {J.}~\bibnamefont
  {Ib{\'a}{\~n}ez}},\ }\href@noop {} {\bibinfo  {journal} {private
  information}\ }\BibitemShut {NoStop}%
\bibitem [{\citenamefont {Graf}\ \emph {et~al.}(2011)\citenamefont {Graf},
  \citenamefont {Stillwell}, \citenamefont {Purcell},\ and\ \citenamefont
  {Tozer}}]{Graf2011}%
  \BibitemOpen
\bibfield  {journal} {  }\bibfield  {author} {\bibinfo {author} {\bibfnamefont
  {D.~E.}\ \bibnamefont {Graf}}, \bibinfo {author} {\bibfnamefont {R.~L.}\
  \bibnamefont {Stillwell}}, \bibinfo {author} {\bibfnamefont {K.~M.}\
  \bibnamefont {Purcell}},\ and\ \bibinfo {author} {\bibfnamefont {S.~W.}\
  \bibnamefont {Tozer}},\ }\href {https://doi.org/10.1080/08957959.2011.633909}
  {\bibfield  {journal} {\bibinfo  {journal} {High Pressure Research}\ }\textbf
  {\bibinfo {volume} {31}},\ \bibinfo {pages} {533} (\bibinfo {year}
  {2011})}\BibitemShut {NoStop}%
\bibitem [{\citenamefont {Kenichi}(2007)}]{Takemura2007}%
  \BibitemOpen
  \bibfield  {author} {\bibinfo {author} {\bibfnamefont {T.}~\bibnamefont
  {Kenichi}},\ }\href {https://doi.org/10.1143/jpsjs.76sa.202} {\bibfield
  {journal} {\bibinfo  {journal} {Journal of the Physical Society of Japan}\
  }\textbf {\bibinfo {volume} {76}},\ \bibinfo {pages} {202} (\bibinfo {year}
  {2007})}\BibitemShut {NoStop}%
\bibitem [{\citenamefont {Giannozzi}\ \emph {et~al.}(2009)\citenamefont
  {Giannozzi}, \citenamefont {Baroni}, \citenamefont {Bonini}, \citenamefont
  {Calandra}, \citenamefont {Car}, \citenamefont {Cavazzoni}, \citenamefont
  {Ceresoli}, \citenamefont {Chiarotti}, \citenamefont {Cococcioni},
  \citenamefont {Dabo}, \citenamefont {Corso}, \citenamefont {de~Gironcoli},
  \citenamefont {Fabris}, \citenamefont {Fratesi}, \citenamefont {Gebauer},
  \citenamefont {Gerstmann}, \citenamefont {Gougoussis}, \citenamefont
  {Kokalj}, \citenamefont {Lazzeri}, \citenamefont {Martin-Samos},
  \citenamefont {Marzari}, \citenamefont {Mauri}, \citenamefont {Mazzarello},
  \citenamefont {Paolini}, \citenamefont {Pasquarello}, \citenamefont
  {Paulatto}, \citenamefont {Sbraccia}, \citenamefont {Scandolo}, \citenamefont
  {Sclauzero}, \citenamefont {Seitsonen}, \citenamefont {Smogunov},
  \citenamefont {Umari},\ and\ \citenamefont {Wentzcovitch}}]{Giannozzi_2009}%
  \BibitemOpen
  \bibfield  {author} {\bibinfo {author} {\bibfnamefont {P.}~\bibnamefont
  {Giannozzi}}, \bibinfo {author} {\bibfnamefont {S.}~\bibnamefont {Baroni}},
  \bibinfo {author} {\bibfnamefont {N.}~\bibnamefont {Bonini}}, \bibinfo
  {author} {\bibfnamefont {M.}~\bibnamefont {Calandra}}, \bibinfo {author}
  {\bibfnamefont {R.}~\bibnamefont {Car}}, \bibinfo {author} {\bibfnamefont
  {C.}~\bibnamefont {Cavazzoni}}, \bibinfo {author} {\bibfnamefont
  {D.}~\bibnamefont {Ceresoli}}, \bibinfo {author} {\bibfnamefont {G.~L.}\
  \bibnamefont {Chiarotti}}, \bibinfo {author} {\bibfnamefont {M.}~\bibnamefont
  {Cococcioni}}, \bibinfo {author} {\bibfnamefont {I.}~\bibnamefont {Dabo}},
  \bibinfo {author} {\bibfnamefont {A.~D.}\ \bibnamefont {Corso}}, \bibinfo
  {author} {\bibfnamefont {S.}~\bibnamefont {de~Gironcoli}}, \bibinfo {author}
  {\bibfnamefont {S.}~\bibnamefont {Fabris}}, \bibinfo {author} {\bibfnamefont
  {G.}~\bibnamefont {Fratesi}}, \bibinfo {author} {\bibfnamefont
  {R.}~\bibnamefont {Gebauer}}, \bibinfo {author} {\bibfnamefont
  {U.}~\bibnamefont {Gerstmann}}, \bibinfo {author} {\bibfnamefont
  {C.}~\bibnamefont {Gougoussis}}, \bibinfo {author} {\bibfnamefont
  {A.}~\bibnamefont {Kokalj}}, \bibinfo {author} {\bibfnamefont
  {M.}~\bibnamefont {Lazzeri}}, \bibinfo {author} {\bibfnamefont
  {L.}~\bibnamefont {Martin-Samos}}, \bibinfo {author} {\bibfnamefont
  {N.}~\bibnamefont {Marzari}}, \bibinfo {author} {\bibfnamefont
  {F.}~\bibnamefont {Mauri}}, \bibinfo {author} {\bibfnamefont
  {R.}~\bibnamefont {Mazzarello}}, \bibinfo {author} {\bibfnamefont
  {S.}~\bibnamefont {Paolini}}, \bibinfo {author} {\bibfnamefont
  {A.}~\bibnamefont {Pasquarello}}, \bibinfo {author} {\bibfnamefont
  {L.}~\bibnamefont {Paulatto}}, \bibinfo {author} {\bibfnamefont
  {C.}~\bibnamefont {Sbraccia}}, \bibinfo {author} {\bibfnamefont
  {S.}~\bibnamefont {Scandolo}}, \bibinfo {author} {\bibfnamefont
  {G.}~\bibnamefont {Sclauzero}}, \bibinfo {author} {\bibfnamefont {A.~P.}\
  \bibnamefont {Seitsonen}}, \bibinfo {author} {\bibfnamefont {A.}~\bibnamefont
  {Smogunov}}, \bibinfo {author} {\bibfnamefont {P.}~\bibnamefont {Umari}},\
  and\ \bibinfo {author} {\bibfnamefont {R.~M.}\ \bibnamefont {Wentzcovitch}},\
  }\href {https://doi.org/10.1088/0953-8984/21/39/395502} {\bibfield  {journal}
  {\bibinfo  {journal} {Journal of Physics: Condensed Matter}\ }\textbf
  {\bibinfo {volume} {21}},\ \bibinfo {pages} {395502} (\bibinfo {year}
  {2009})}\BibitemShut {NoStop}%
\bibitem [{\citenamefont {Giannozzi}\ \emph {et~al.}(2017)\citenamefont
  {Giannozzi}, \citenamefont {Andreussi}, \citenamefont {Brumme}, \citenamefont
  {Bunau}, \citenamefont {Nardelli}, \citenamefont {Calandra}, \citenamefont
  {Car}, \citenamefont {Cavazzoni}, \citenamefont {Ceresoli}, \citenamefont
  {Cococcioni}, \citenamefont {Colonna}, \citenamefont {Carnimeo},
  \citenamefont {Corso}, \citenamefont {de~Gironcoli}, \citenamefont {Delugas},
  \citenamefont {DiStasio}, \citenamefont {Ferretti}, \citenamefont {Floris},
  \citenamefont {Fratesi}, \citenamefont {Fugallo}, \citenamefont {Gebauer},
  \citenamefont {Gerstmann}, \citenamefont {Giustino}, \citenamefont {Gorni},
  \citenamefont {Jia}, \citenamefont {Kawamura}, \citenamefont {Ko},
  \citenamefont {Kokalj}, \citenamefont {Küçükbenli}, \citenamefont
  {Lazzeri}, \citenamefont {Marsili}, \citenamefont {Marzari}, \citenamefont
  {Mauri}, \citenamefont {Nguyen}, \citenamefont {Nguyen}, \citenamefont {de-la
  Roza}, \citenamefont {Paulatto}, \citenamefont {Poncé}, \citenamefont
  {Rocca}, \citenamefont {Sabatini}, \citenamefont {Santra}, \citenamefont
  {Schlipf}, \citenamefont {Seitsonen}, \citenamefont {Smogunov}, \citenamefont
  {Timrov}, \citenamefont {Thonhauser}, \citenamefont {Umari}, \citenamefont
  {Vast}, \citenamefont {Wu},\ and\ \citenamefont {Baroni}}]{Giannozzi_2017}%
  \BibitemOpen
  \bibfield  {author} {\bibinfo {author} {\bibfnamefont {P.}~\bibnamefont
  {Giannozzi}}, \bibinfo {author} {\bibfnamefont {O.}~\bibnamefont
  {Andreussi}}, \bibinfo {author} {\bibfnamefont {T.}~\bibnamefont {Brumme}},
  \bibinfo {author} {\bibfnamefont {O.}~\bibnamefont {Bunau}}, \bibinfo
  {author} {\bibfnamefont {M.~B.}\ \bibnamefont {Nardelli}}, \bibinfo {author}
  {\bibfnamefont {M.}~\bibnamefont {Calandra}}, \bibinfo {author}
  {\bibfnamefont {R.}~\bibnamefont {Car}}, \bibinfo {author} {\bibfnamefont
  {C.}~\bibnamefont {Cavazzoni}}, \bibinfo {author} {\bibfnamefont
  {D.}~\bibnamefont {Ceresoli}}, \bibinfo {author} {\bibfnamefont
  {M.}~\bibnamefont {Cococcioni}}, \bibinfo {author} {\bibfnamefont
  {N.}~\bibnamefont {Colonna}}, \bibinfo {author} {\bibfnamefont
  {I.}~\bibnamefont {Carnimeo}}, \bibinfo {author} {\bibfnamefont {A.~D.}\
  \bibnamefont {Corso}}, \bibinfo {author} {\bibfnamefont {S.}~\bibnamefont
  {de~Gironcoli}}, \bibinfo {author} {\bibfnamefont {P.}~\bibnamefont
  {Delugas}}, \bibinfo {author} {\bibfnamefont {R.~A.}\ \bibnamefont
  {DiStasio}}, \bibinfo {author} {\bibfnamefont {A.}~\bibnamefont {Ferretti}},
  \bibinfo {author} {\bibfnamefont {A.}~\bibnamefont {Floris}}, \bibinfo
  {author} {\bibfnamefont {G.}~\bibnamefont {Fratesi}}, \bibinfo {author}
  {\bibfnamefont {G.}~\bibnamefont {Fugallo}}, \bibinfo {author} {\bibfnamefont
  {R.}~\bibnamefont {Gebauer}}, \bibinfo {author} {\bibfnamefont
  {U.}~\bibnamefont {Gerstmann}}, \bibinfo {author} {\bibfnamefont
  {F.}~\bibnamefont {Giustino}}, \bibinfo {author} {\bibfnamefont
  {T.}~\bibnamefont {Gorni}}, \bibinfo {author} {\bibfnamefont
  {J.}~\bibnamefont {Jia}}, \bibinfo {author} {\bibfnamefont {M.}~\bibnamefont
  {Kawamura}}, \bibinfo {author} {\bibfnamefont {H.-Y.}\ \bibnamefont {Ko}},
  \bibinfo {author} {\bibfnamefont {A.}~\bibnamefont {Kokalj}}, \bibinfo
  {author} {\bibfnamefont {E.}~\bibnamefont {Küçükbenli}}, \bibinfo {author}
  {\bibfnamefont {M.}~\bibnamefont {Lazzeri}}, \bibinfo {author} {\bibfnamefont
  {M.}~\bibnamefont {Marsili}}, \bibinfo {author} {\bibfnamefont
  {N.}~\bibnamefont {Marzari}}, \bibinfo {author} {\bibfnamefont
  {F.}~\bibnamefont {Mauri}}, \bibinfo {author} {\bibfnamefont {N.~L.}\
  \bibnamefont {Nguyen}}, \bibinfo {author} {\bibfnamefont {H.-V.}\
  \bibnamefont {Nguyen}}, \bibinfo {author} {\bibfnamefont {A.~O.}\
  \bibnamefont {de-la Roza}}, \bibinfo {author} {\bibfnamefont
  {L.}~\bibnamefont {Paulatto}}, \bibinfo {author} {\bibfnamefont
  {S.}~\bibnamefont {Poncé}}, \bibinfo {author} {\bibfnamefont
  {D.}~\bibnamefont {Rocca}}, \bibinfo {author} {\bibfnamefont
  {R.}~\bibnamefont {Sabatini}}, \bibinfo {author} {\bibfnamefont
  {B.}~\bibnamefont {Santra}}, \bibinfo {author} {\bibfnamefont
  {M.}~\bibnamefont {Schlipf}}, \bibinfo {author} {\bibfnamefont {A.~P.}\
  \bibnamefont {Seitsonen}}, \bibinfo {author} {\bibfnamefont {A.}~\bibnamefont
  {Smogunov}}, \bibinfo {author} {\bibfnamefont {I.}~\bibnamefont {Timrov}},
  \bibinfo {author} {\bibfnamefont {T.}~\bibnamefont {Thonhauser}}, \bibinfo
  {author} {\bibfnamefont {P.}~\bibnamefont {Umari}}, \bibinfo {author}
  {\bibfnamefont {N.}~\bibnamefont {Vast}}, \bibinfo {author} {\bibfnamefont
  {X.}~\bibnamefont {Wu}},\ and\ \bibinfo {author} {\bibfnamefont
  {S.}~\bibnamefont {Baroni}},\ }\href
  {https://doi.org/10.1088/1361-648X/aa8f79} {\bibfield  {journal} {\bibinfo
  {journal} {Journal of Physics: Condensed Matter}\ }\textbf {\bibinfo {volume}
  {29}},\ \bibinfo {pages} {465901} (\bibinfo {year} {2017})}\BibitemShut
  {NoStop}%
\end{thebibliography}%


\begin{thebibliography}{4}%
\makeatletter
\providecommand \@ifxundefined [1]{%
 \@ifx{#1\undefined}
}%
\providecommand \@ifnum [1]{%
 \ifnum #1\expandafter \@firstoftwo
 \else \expandafter \@secondoftwo
 \fi
}%
\providecommand \@ifx [1]{%
 \ifx #1\expandafter \@firstoftwo
 \else \expandafter \@secondoftwo
 \fi
}%
\providecommand \natexlab [1]{#1}%
\providecommand \enquote  [1]{``#1''}%
\providecommand \bibnamefont  [1]{#1}%
\providecommand \bibfnamefont [1]{#1}%
\providecommand \citenamefont [1]{#1}%
\providecommand \href@noop [0]{\@secondoftwo}%
\providecommand \href [0]{\begingroup \@sanitize@url \@href}%
\providecommand \@href[1]{\@@startlink{#1}\@@href}%
\providecommand \@@href[1]{\endgroup#1\@@endlink}%
\providecommand \@sanitize@url [0]{\catcode `\\12\catcode `\$12\catcode `\&12\catcode `\#12\catcode `\^12\catcode `\_12\catcode `\%12\relax}%
\providecommand \@@startlink[1]{}%
\providecommand \@@endlink[0]{}%
\providecommand \url  [0]{\begingroup\@sanitize@url \@url }%
\providecommand \@url [1]{\endgroup\@href {#1}{\urlprefix }}%
\providecommand \urlprefix  [0]{URL }%
\providecommand \Eprint [0]{\href }%
\providecommand \doibase [0]{https://doi.org/}%
\providecommand \selectlanguage [0]{\@gobble}%
\providecommand \bibinfo  [0]{\@secondoftwo}%
\providecommand \bibfield  [0]{\@secondoftwo}%
\providecommand \translation [1]{[#1]}%
\providecommand \BibitemOpen [0]{}%
\providecommand \bibitemStop [0]{}%
\providecommand \bibitemNoStop [0]{.\EOS\space}%
\providecommand \EOS [0]{\spacefactor3000\relax}%
\providecommand \BibitemShut  [1]{\csname bibitem#1\endcsname}%
\let\auto@bib@innerbib\@empty
\bibitem [{\citenamefont {Kenichi}(2007)}]{takemura2007}%
  \BibitemOpen
  \bibfield  {author} {\bibinfo {author} {\bibfnamefont {T.}~\bibnamefont {Kenichi}},\ }\href {https://doi.org/10.1143/jpsjs.76sa.202} {\bibfield  {journal} {\bibinfo  {journal} {Journal of the Physical Society of Japan}\ }\textbf {\bibinfo {volume} {76}},\ \bibinfo {pages} {202} (\bibinfo {year} {2007})}\BibitemShut {NoStop}%
\bibitem [{\citenamefont {Klotz}\ \emph {et~al.}(2009)\citenamefont {Klotz}, \citenamefont {Chervin}, \citenamefont {Munsch},\ and\ \citenamefont {{Le Marchand}}}]{Klotz2009}%
  \BibitemOpen
  \bibfield  {author} {\bibinfo {author} {\bibfnamefont {S.}~\bibnamefont {Klotz}}, \bibinfo {author} {\bibfnamefont {J.~C.}\ \bibnamefont {Chervin}}, \bibinfo {author} {\bibfnamefont {P.}~\bibnamefont {Munsch}},\ and\ \bibinfo {author} {\bibfnamefont {G.}~\bibnamefont {{Le Marchand}}},\ }\bibfield  {journal} {\bibinfo  {journal} {Journal of Physics D: Applied Physics}\ }\textbf {\bibinfo {volume} {42}},\ \href {https://doi.org/10.1088/0022-3727/42/7/075413} {10.1088/0022-3727/42/7/075413} (\bibinfo {year} {2009})\BibitemShut {NoStop}%
\bibitem [{\citenamefont {Cardona}\ and\ \citenamefont {Guntherodt}(1982)}]{Cardona1982}%
  \BibitemOpen
  \bibfield  {author} {\bibinfo {author} {\bibfnamefont {M.}~\bibnamefont {Cardona}}\ and\ \bibinfo {author} {\bibfnamefont {G.}~\bibnamefont {Guntherodt}},\ }\href@noop {} {\bibfield  {journal} {\bibinfo  {journal} {Topics in Applied Physics}\ }\textbf {\bibinfo {volume} {50}} (\bibinfo {year} {1982})}\BibitemShut {NoStop}%
\bibitem [{\citenamefont {Pimenta}\ \emph {et~al.}(2021)\citenamefont {Pimenta}, \citenamefont {Resende}, \citenamefont {Ribeiro},\ and\ \citenamefont {Carvalho}}]{Pimenta2021}%
  \BibitemOpen
  \bibfield  {author} {\bibinfo {author} {\bibfnamefont {M.~A.}\ \bibnamefont {Pimenta}}, \bibinfo {author} {\bibfnamefont {G.~C.}\ \bibnamefont {Resende}}, \bibinfo {author} {\bibfnamefont {H.~B.}\ \bibnamefont {Ribeiro}},\ and\ \bibinfo {author} {\bibfnamefont {B.~R.}\ \bibnamefont {Carvalho}},\ }\href {https://doi.org/10.1039/D1CP03626B} {\bibfield  {journal} {\bibinfo  {journal} {Phys. Chem. Chem. Phys.}\ }\textbf {\bibinfo {volume} {23}},\ \bibinfo {pages} {27103} (\bibinfo {year} {2021})}\BibitemShut {NoStop}%
\end{thebibliography}%
\end{document}


\newcommand{\HfS}{$\text{HfS}_{2}$\xspace}

\title{Pressure-induced optical anisotropy of \HfS}


\def \FUW{Faculty of Physics, University
of Warsaw, Pasteura 5, 02-093 Warsaw, Poland}
\def \LNCMI{Laboratoire National des Champs Magnétiques Intenses, CNRS-UGA-UPS-INSA-EMFL, Grenoble, France}
\def \China{Hefei Innovation Research Institute, School of Microelectronics, Beihang University, Hefei 230013, P. R. China}
\def \CENT{Centre of New Technologies, University of Warsaw, Banacha 2c, 02-097 Warsaw, Poland}
\def \Wroclaw{Department of Semiconductor Materials Engineering, Faculty of Fundamental Problems of Technology, Wrocław University of Science and Technology, Wybrzeże Wyspiańskiego 27, 50-370, Wrocław, Poland}
\def \Spain{Geosciences Barcelona (GEO3BCN), CSIC, Lluís Solé i Sabarís s.n., 08028, Barcelona, Catalonia, Spain}
\def \MagdaG{Department of Materials Science and Engineering, National University of Singapore, 117575, Singapore} 
\def \MagdaGk{Institute for Functional Intelligent Materials, National University of Singapore, 117544, Singapore}

\author{Igor Antoniazzi} 
\email{igor.antoniazzi@fuw.edu.pl}
\affiliation{\FUW}
\author{Tomasz Woźniak}
\affiliation{\FUW}
\author{Amit Pawbake}
\affiliation{\LNCMI}
\author{Natalia Zawadzka}
\affiliation{\FUW}
\author{Magdalena Grzeszczyk}
\affiliation{\FUW}
\affiliation{\MagdaGk}
\author{Zahir Muhammad}
\affiliation{\China}
\author{Weisheng Zhao}
\affiliation{\China}
\author{Jordi~Ib{\'a}{\~n}ez}
\affiliation{\Spain}
\author{Clement Faugeras}
\affiliation{\LNCMI}
\author{Maciej R. Molas}
\affiliation{\FUW}
\author{Adam Babiński}\email{adam.babinski@fuw.edu.pl}
\affiliation{\FUW}

\begin{abstract} 

\end{abstract}

\maketitle

\section{Supplementary Material}

\subsection{Angle-resolved polarized Raman scattering under pressure}

\begin{figure}[h]
    \centering
    \includegraphics[width=.6\linewidth]{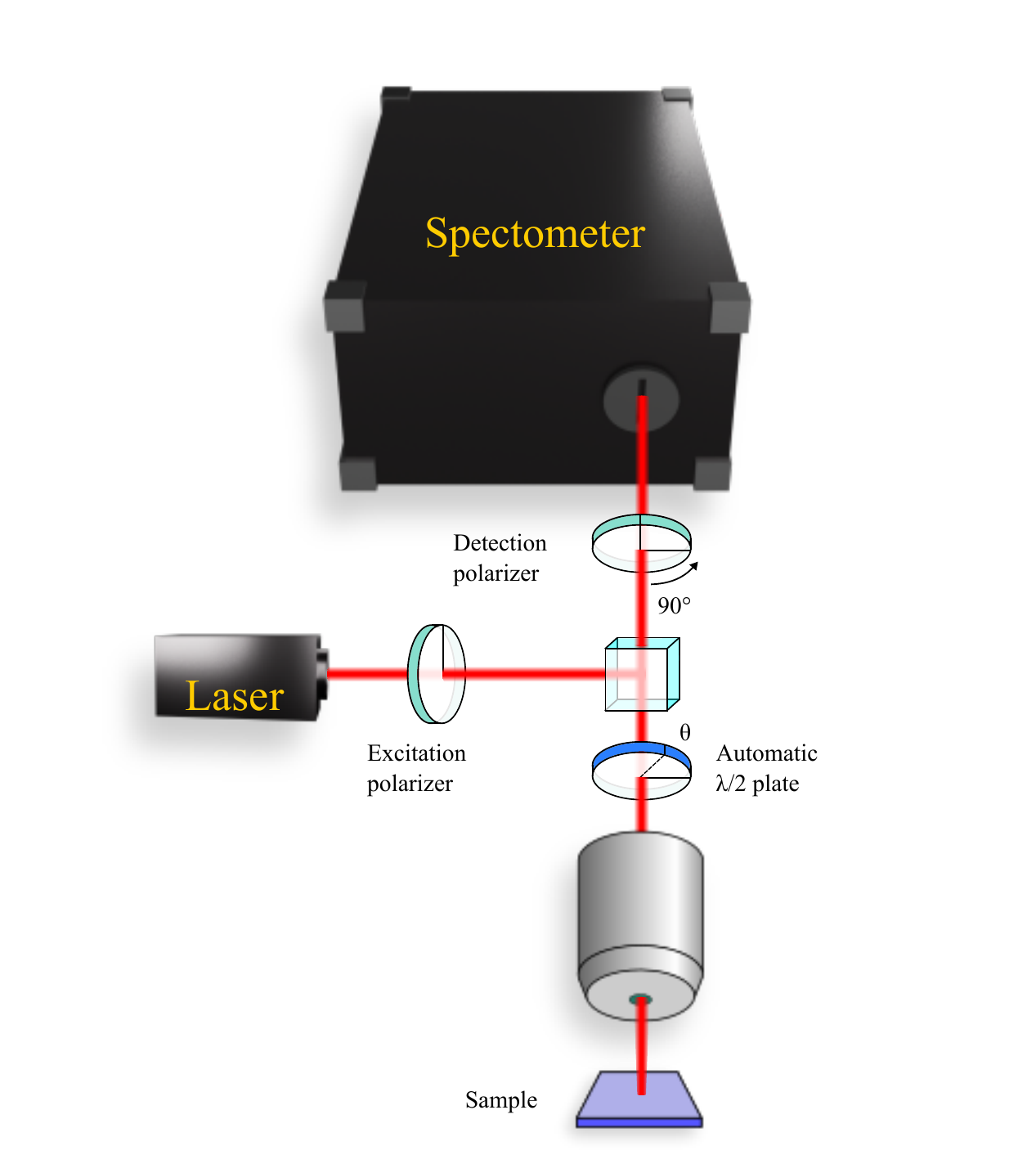}
    \caption{\label{si: fig. s2}
    Schematic illustration of the experimental setup used for ARPR measurements.
    }
\end{figure}

A schematic illustration of the experimental setup used for the measurements of the angle-resolved polarized Raman scattering (ARPR) is shown in Fig.~\ref{si: fig. s2}.
There are two linear polarizers and a half-wave plate in the optical path. 
One polarizer is placed in the excitation path, directly after the laser output, while the second one is placed in the detection path, just before the spectrometer. 
The relative orientation of polarization axes of the excitation and detection polarizers permits to choose between co- (XX) and cross-linear (XY) configurations of the experiment. 
The rotation of a half-wave plate, mounted right on top of the objective, allows to rotate simultaneously both the excitation and detection polarizations in respect of the studied sample.
The presented setup is analog of the rotation of the sample.

\clearpage

\subsection{\HfS Raman tensors}
The \HfS Raman tensors expected under ambient pressure for $\textrm{A}_\textrm{1g}$ and $\textrm{E}_\textrm{g}$ (degenerated) are:

\begin{equation}
\centering
\textrm{A}_\textrm{1g}=\begin{pmatrix} a & 0 & 0 \\ 0 & a & 0 \\ 0 & 0 & b \end{pmatrix};
\textrm{E}_\textrm{g,1}=\begin{pmatrix} c & 0 & 0 \\ 0 & -c & d \\ 0 & d & 0 \end{pmatrix},
\textrm{E}_\textrm{g,2}=\begin{pmatrix} 0 & -c & -d \\ -c & 0 & 0 \\ -d & 0 & 0 \end{pmatrix}
\label{eqs: HfS_Tensor1}
\end{equation}

The calculated polar intensity dependence for both configuration XX and XY are:

\vspace{-10 pt}
\begin{align}
   &\textrm{A}_\textrm{1g}\textrm{(XX)}= a^2 \label{Eq: 3} \\
   &\textrm{A}_\textrm{1g}\textrm{(XY)}=0 \label{Eq: 4}\\
   &\textrm{E}_\textrm{g}\textrm{(XX)}= c^2 \\
   &\textrm{E}_\textrm{g}\textrm{(XY)}= c^2  
\end{align}

\clearpage

\subsection{Experimental conditions of the experiment}

\begin{figure}[h]
    \centering
    \includegraphics[width=1\linewidth]{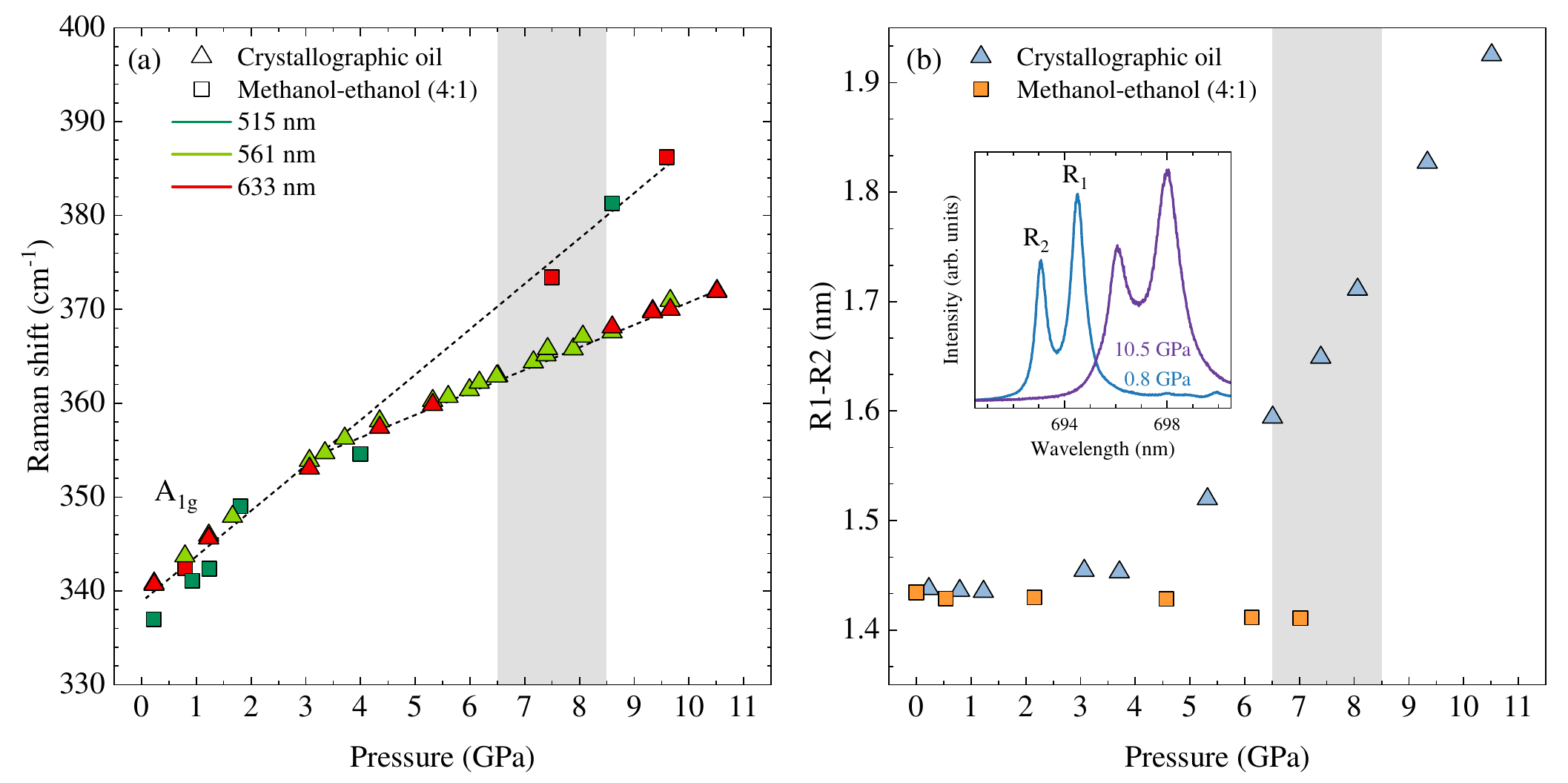}
    \caption{\label{si: fig. s1} 
    (a) Pressure dependence of the A$_\textrm{1g}$ frequencies to three different excitation energies, 515 nm (2.41 eV), 561 nm (2.21 eV, and 633 nm (1.96 eV) and two PTM, a crystallographic oil based on Silicone and the traditional methanol-ethanol (4:1).
    (b) The uniaxial stress as a function of pressure for crystallographic oil and methanol-ethanol.
    }
\end{figure}

Crucial for our conclusions is the non-hydrostaticity of the pressure transmitting medium (PTM).~\cite{takemura2007, Klotz2009}
In Fig.~\ref{si: fig. s1}.(a) we present the pressure dependence of the A$_\textrm{1g}$ frequencies to three different excitation energies, 515 nm, 561 nm, and 633 nm and two PTM, a crystallographic oil based on Silicone and the traditional methanol-ethanol (4:1).
The experiment was sensitive only to the different PTM.
The pressure was measured by the Ruby line $R_1$ shifting.
In Fig.~\ref{si: fig. s1}.(b) we present the called uniaxial stress, which can be measured by the separation $R_1-R_2$, for both PTM.
To $P$>5 GPa the uniaxial stress to crystallographic oil increases as far as pressure goes further.
Showing the non-hydrostaticity of the PTM for such high pressure (HP).
In the inset is presented the Ruby fluorescence under 0.8 GPa and 10.5 GPa.

\clearpage

\subsection{Pressure evolution under 561 nm}
All the features discussed in the main text were observed for both $\lambdaup$=561 nm and 633 nm excitations.
Slightly differences in the shape spectra are present mainly in the pressure range 7.3 GPa<$P$<9.4 GPa.

\begin{figure}[h]
    \centering
    \includegraphics[width=0.7\linewidth]{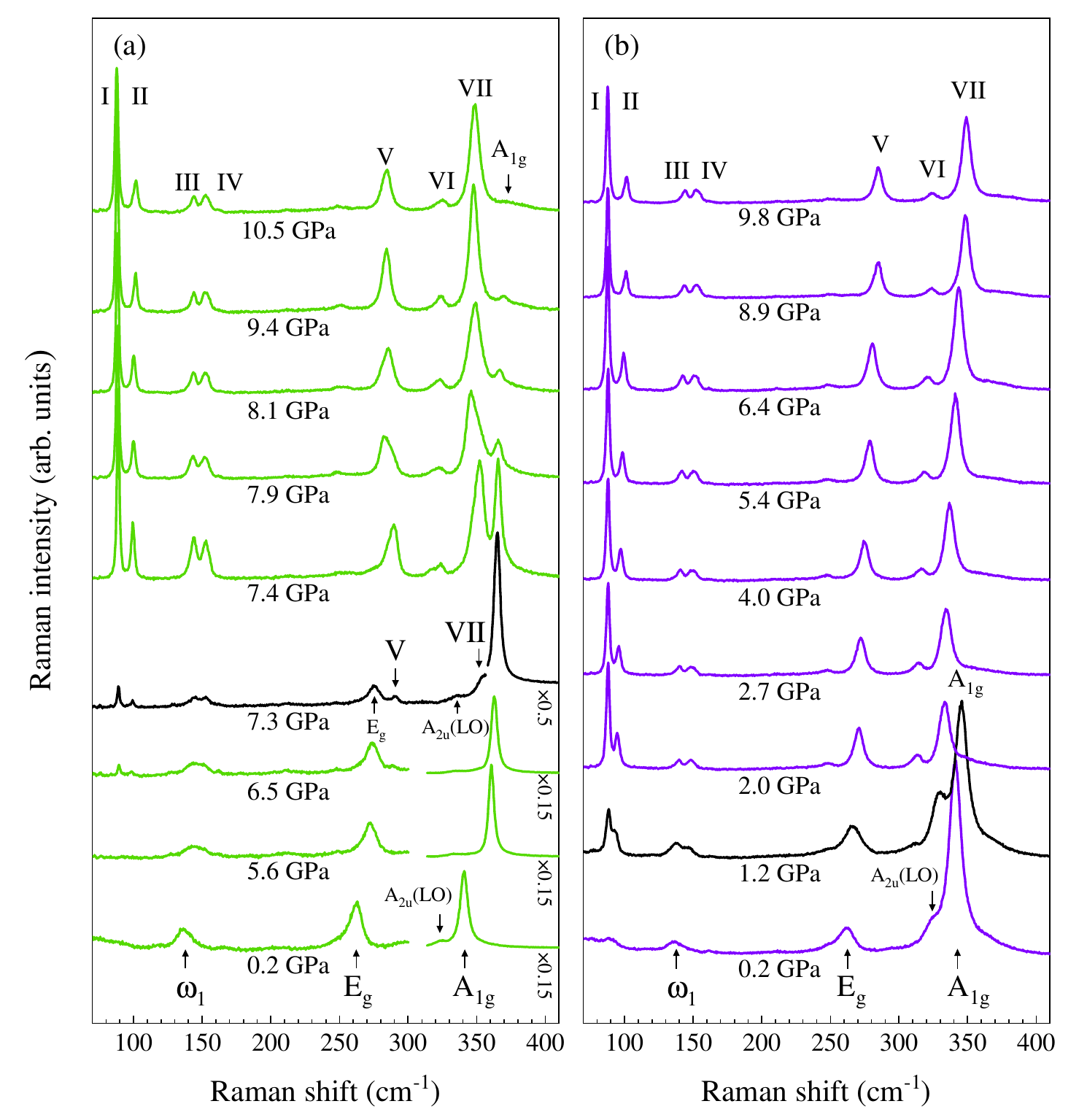}
    \caption{\label{si: HP_RS_633nm} 
    The evolution of RS in bulk HfS$_2$ during compression (a) and decompression (b). The measurements were taken under $\lambdaup$=561 nm excatation.
    }
\end{figure}

\clearpage

\subsection{Pressure fitting to 7.3 GPa<$P$<9.4 GPa}
\begin{figure}[h]
    \centering
    \includegraphics[width=.9\linewidth]{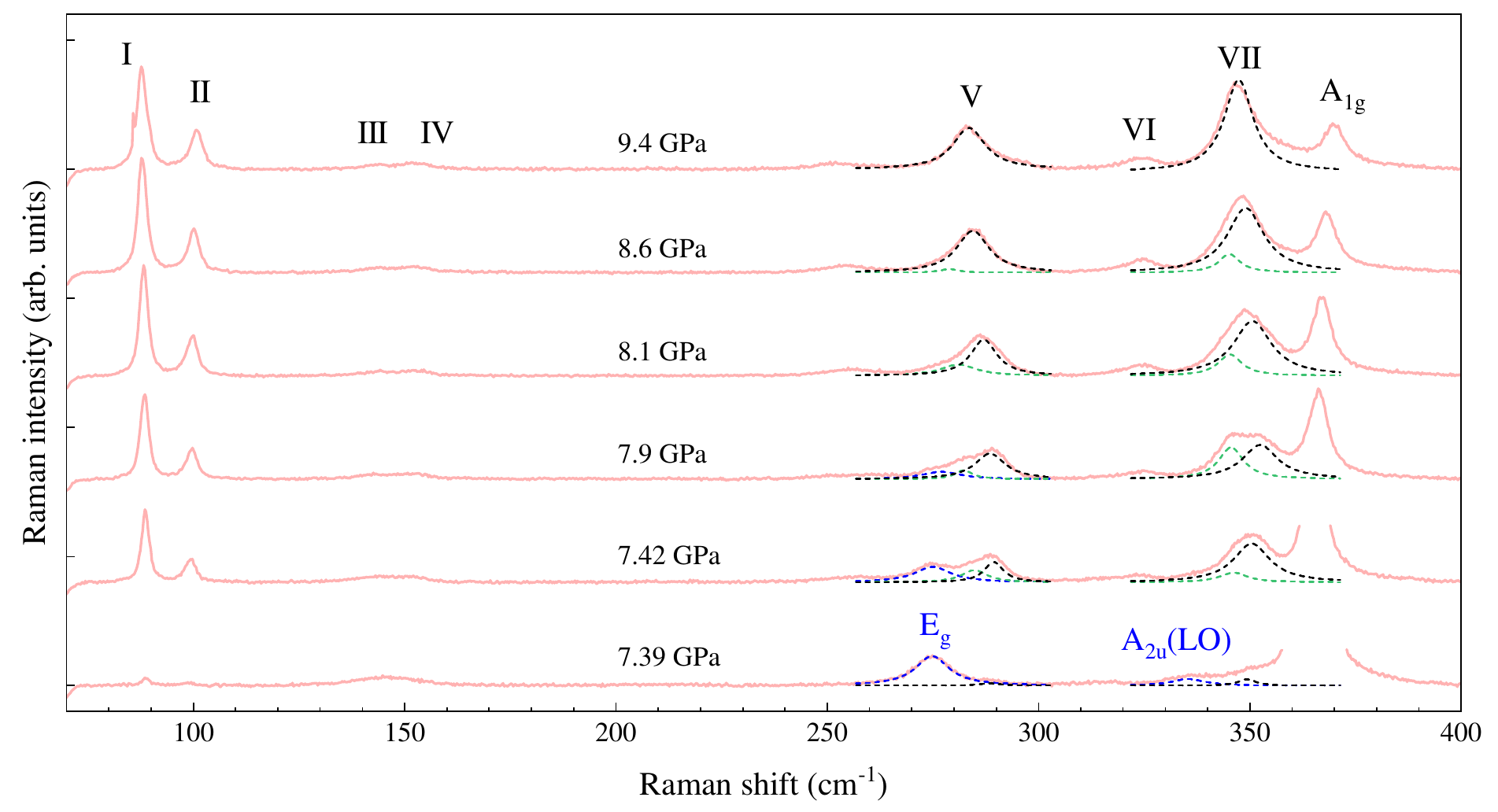}
    \caption{\label{si: fig. s3} 
    RS pressure evolution in the range 7.3 GPa<$P$<9.4 GPa.
    The dashed lines represents the three different components: blue the ambient modes A$_\textrm{2u}$(LO) and E$_\textrm{g}$, black the HP peaks VII and V and green the extra components.
    The measurements were taken under excitation 633 nm.
    }
\end{figure}

As discussed in the main text, the emergence of VII and V takes place together the appearance of an extra mode to each one.
The Fig.~\ref{si: fig. s3} presents the RS pressure evolution in the range 7.3 GPa<$P$<9.4 GPa highlighting the importance of the mentioned extra modes.
These extra components got stronger to $P$=7.9 GPa, even though changing the spectrum shape.
As the pressure goes further their intensity decreased, reaching zero under $P$=9.4 GPa.

\clearpage

\subsection{Raman frequencies pressure dependence}
The whole pressure evolution parameters calculated to compression and decompression are presented in the Table.~\ref{Tbs: SI}.

\begin{table}[h]
\centering
 \begin{subtable}[t]{0.45\linewidth}
    \centering
      \begin{tabular}[t]{c|c|c}
      \multicolumn{3}{c}{Parameters compression}\\
      & $E$ (cm$^{-1}$) & $\alpha$(cm$^{-1}$/GPa) \\
      \hline \hline
      $\omegaup_1$  & 136.3 & 1.36 \\
      E$_\textrm{g}$  & 261.5 & 1.92 \\
      A$_\textrm{2u}$(LO) & 322.4 & 2.13\\
      A$_\textrm{1g}$ ($P$<5.5 GPa) & 340.7 & 3.84\\ 
      A$_\textrm{1g}$ ($P$>5.5 GPa) & 347.8 & 2.33 \\ 
      I & 92.0 & -0.44 \\
      II & 93.0 & 0.82 \\
      III & 141.1 & 0.21 \\
      IV & 144.3 & 0.87 \\
      V & - & - \\
      VI & - & - \\
      VII & - & - \\
      \end{tabular}
 \end{subtable}%
 \begin{subtable}[t]{0.45\linewidth}
    \centering
      \begin{tabular}[t]{c|c|c}
      \multicolumn{3}{c}{Parameters decompression}\\
      & $E$ (cm$^{-1}$) & $\alpha$(cm$^{-1}$/GPa) \\
      \hline \hline
      $\omegaup_1$  & - & - \\
      E$_\textrm{g}$  & - & - \\
      A$_\textrm{2u}$(LO) & - & - \\
      A$_\textrm{1g}$ ($P$<5.5 GPa) & 340.9 & 3.84\\ 
      A$_\textrm{1g}$ ($P$>5.5 GPa) & 351.7 & 2.04 \\
      I & 88.3 & -0.07 \\
      II & 92.9 & 0.90 \\
      III & 137.2 & 0.69 \\
      IV (P>2.0 GPa)& 145.7 & 0.85 \\
      V (P>2.0 GPa)& 268.5 & 1.70 \\
      VI (P>2.0 GPa)& 310.4 & 1.50 \\
      VII (P>2.0 GPa) & 330.1 & 2.01 \\
      \end{tabular}
 \end{subtable}
 \caption{
    Pressure-dependent parameters calculated to the compression and decompression.
    The measurements were taken under $\lambdaup$=633 nm excitation.
    }
\label{Tbs: SI}    
\end{table}

\clearpage

\subsection{ARPRS under $P$=1.2 GPa}

\begin{figure}[hb!]
    \centering
    \includegraphics[width=1\linewidth]{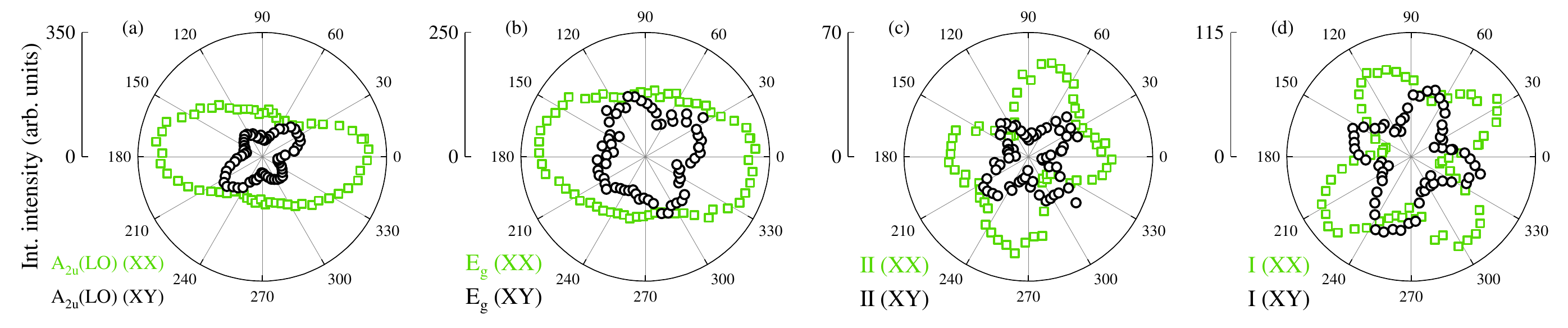}
    \caption{\label{si:Set_LPpol}    
    The decompressed angular dependence to A$_\textrm{2u}$(LO) (a), E$_\textrm{g}$ (b), II (c), I (d) under $P$=1.2 GPa.
    The colorful open squares and circles represents the XX and XY configuration, respectively.
    The measurements were taken under wavelength $\lambdaup$=561 nm with 500 $\muup$W.
    }
\end{figure}

The decompressed angle resolved polarized Raman (ARPR) under $P$=1.2 GPa can be appreciated in the Fig.~\ref{si:Set_LPpol} for A$_\textrm{2u}$(LO) (b, c), II (d,e), I (f, g), and E$_\textrm{g}$ (h, i).
To this pressure we consider that VII and V turned back into A$_\textrm{2u}$(LO) and E$_\textrm{g}$, respectively.

\clearpage
\subsection{ARPRS under pressure details}

To fit the experimental data we considered several approaches to the Raman tensor.
The Raman tensor for $\textrm{A}_\textrm{1g}$ in 1T-\HfS takes the form presented in Eq.~\ref{eqs: R_Tensor1}.

\begin{equation}
\centering
\textrm{R}=\begin{pmatrix} a & 0 \\ 0 & a  \end{pmatrix}
\label{eqs: R_Tensor1}
\end{equation}

This leads to a symmetric polar dependence of the RS in the XX configuration and to the quenching of the mode in the XY configuration:

\vspace{-10 pt}
\begin{align}
   &\textrm{I(XX)}= a^2 \label{R1(XX)} \\
   &\textrm{I(XY)}= 0 \label{R1(XY)}
\end{align}

The Raman tensor~\cite{Cardona1982} which corresponds to A/A$_1$/A$_\textrm{g}$ Raman-active modes in a crystal of an orthorhombic symmetry~\cite{Pimenta2021} is shown in Eq~\ref{eqs: R_Tensor2}:

\begin{equation}
\centering
\textrm{R}=\begin{pmatrix} ae^{i\varphi_a} & 0 \\ 0 & be^{i\varphi_b}  \end{pmatrix} = ae^{i\varphi_a}\begin{pmatrix} 1 & 0 \\ 0 & (b/a)e^{i\varphi_{ba}}  \end{pmatrix}
\label{eqs: R_Tensor2}
\end{equation}
\noindent where $\varphi_{ij}=\varphi_i-\varphi_j$.
The polar dependence of the RS intensity expected for the Raman tensor for the A/A$_1$/A$_\textrm{g}$ modes in a crystal of an orthorhombic symmetry in the XX and XY configurations can be expressed as:

\vspace{-10 pt}
\begin{equation}
\textrm{I(XX)}\propto\left[\cos^2\theta+\left(\frac{b}{a}\right)\cos\varphi_{ba}\sin^2\theta\right]^2+\left(\frac{b}{a}\right)^2\sin^4\theta\sin^2\varphi_{ba}
\label{eqs:I(OrthoXX)}
\end{equation}
\begin{equation}
\begin{split}
\textrm{I(XY)}\propto\left[1-\left(\left(\frac{b}{a}\right)\cos\varphi_{ba}\right)^2+\left(\frac{b}{a}\right)^2\sin^2\varphi_{ba}\right]\sin^2\theta\cos^2\theta
\label{eqs:I(OrthoXY)}
\end{split}
\end{equation}
\\

The Raman tensor for Raman-active A$_\textrm{g}$ mode in a crystal of a triclinic symmetry is shown in Eq~\ref{eqs: R_Tensor3}:

\begin{equation}
\centering
\textrm{R}=\begin{pmatrix} ae^{i\varphi_a} & de^{i\varphi_d} \\ de^{i\varphi_d} & be^{i\varphi_b}  \end{pmatrix} = ae^{i\varphi_a}\begin{pmatrix} 1 & (d/a)e^{i\varphi_{da}}\\ (d/a)e^{i\varphi_{da}} & (b/a)e^{i\varphi_{ba}}  \end{pmatrix}
\label{eqs: R_Tensor3}
\end{equation}

The polar dependence of the RS intensity expected for the Raman tensor for the A$_\textrm{g}$ mode in a crystal of a triclinic symmetry in the XX and XY configurations can be expressed as:

\vspace{-10 pt}
\begin{equation}
\begin{split}
   \textrm{I(XX)}\propto\cos^4\theta+\left(\frac{b}{a}\right)^2\sin^4\theta+&\left(\frac{d}{a}\right)^2\sin^22\theta+\sin2\theta\left[\left(\frac{b}{2a}\right)\cos\varphi_{ba}\sin2\theta\right]\\
   -2&\left(\frac{d}{a}\right)\cos\varphi_{da}\cos^2\theta-2\left(\frac{bd}{a^2}\right)\cos(\varphi_{da}-\varphi_{ba})\sin^2\theta 
\label{eqs:I(TriXX)}
\end{split}
\end{equation}

\begin{equation}
\begin{split}
\textrm{I(XY)}\propto\left(\frac{d}{a}\right)^2\cos^22\theta+&\frac{1}{4}\sin^22\theta\left[1-\left(\frac{b}{a}\right)^2-2\left(\frac{b}{a}\right)\cos\varphi_{ba}\right] \\
        -&\frac{1}{2}\sin4\theta\left(\frac{bd}{a^2}\right)\cos(\varphi_{da}-\varphi_{da})-\left(\frac{d}{a}\right)\cos(\varphi_{da})
\label{eqs:I(TriXY)}
\end{split}
\end{equation}
\\

The most general, non-symmetric form the Raman tensor can be expressed aby Eq.~\ref{eqs: R_Tensor4}:

\begin{equation}
\centering
\textrm{R}=\begin{pmatrix} ae^{i\varphi_a} & ce^{i\varphi_c} \\ fe^{i\varphi_f} & be^{i\varphi_b}  \end{pmatrix} = ae^{i\varphi_a}\begin{pmatrix} 1 & (c/a)e^{i\varphi_{ca}}\\ (f/a)e^{i\varphi_{fa}} & (b/a)e^{i\varphi_{ba}}  \end{pmatrix}
\label{eqs: R_Tensor4}
\end{equation}

\noindent The polar dependence of the non-symmetric form the Raman tensor intensity for both polarization configurations can be expressed by:

\begin{equation}
\begin{split}
    I(XX)\propto\;\cos^4\theta+\left(\frac{b}{a} \right)^2\sin^4\theta+\frac{1}{4}\sin^2(2\theta)&\left[ \left( \frac{f}{a} \right)^2+\left( \frac{c}{a} \right)^2+2\frac{fc}{a^2}\cos\varphi_{fc}+2\frac{b}{a}\cos\varphi_{ba}) \right] \\
          \;+2\frac{b}{a}\sin^3\theta\cos\theta&\left[\frac{f}{a}\cos\varphi_{bf}+\frac{c}{a}\cos\varphi_{bc}\right]\\
          \;+2\sin\theta\cos^3\theta&\left[\frac{f}{a}\cos\varphi_{fa}+\frac{c}{a}\cos\varphi_{ca}\right]
\label{eqs:I(XX)}
\end{split}
\end{equation}
\\
\begin{equation}
\begin{split}
    I(XY)\propto\;\left(\frac{f}{a}\right)^2\cos^4\theta+\left(\frac{c}{a}\right)^2\sin^4\theta+\frac{1}{4}\sin^2(2\theta)&\left[1+\left(\frac{b}{a}\right)^2-2\frac{b}{a}\cos\varphi_{ba}-2\frac{cf}{a^2}\cos\varphi_{cf}\right]\\
          \;-2\frac{c}{a}\sin^3\theta\cos\theta&\left[\frac{b}{a}\cos\varphi_{bc}-\cos\varphi_{ca}\right]\\
          \;+2\frac{f}{a}\sin\theta\cos^3\theta&\left[\frac{b}{a}\cos\varphi_{bf}-\cos\varphi_{fa}\right]
\label{eqs:I(XX)}
\end{split}
\end{equation}
\\
\noindent The fittings were performed using the Raman tensor R presented above (Eqs.~\ref{eqs: R_Tensor1},~\ref{eqs: R_Tensor2},~\ref{eqs: R_Tensor3},~\ref{eqs: R_Tensor4}).
The XX configuration was fitted first and the found parameters were used then to emulate the corresponding  XY.

\begin{table}[h]
\centering
\renewcommand{\arraystretch}{1.4}
 \begin{subtable}[t]{0.45\linewidth}
    \centering
     \begin{tabular}[t]{ c|c c|c c| c} 
         Peak & b/a & d/a & $\varphiup_\textrm{ba}$ & $\varphiup_\textrm{da}$ & main axis \\
        \hline \hline
         A$_\textrm{1g}$ & 0.94 & - &103° & - &85° \\
         VII & 0.67 & 0.12 & 270° & 285° & 160° \\
         V & 0.27 & 0.35 & 348° & 103° &170°\\ 
         IV & 0.91 & - & 93° & - & 175°\\ 
         III & 0.97 & - & 90° & - & 165°\\ 
         II & 0.66 & - & 143° & - & 65°\\
         I & 0.77 & 0.05 & 128° & 100° & 35°\\
 
      \end{tabular} 
 \end{subtable}%
 \begin{subtable}[t]{0.45\linewidth}
    \centering
      \begin{tabular}[t]{ c|c c c c|c c c c| c} 
         Peak & b/a & d/a & c/a & f/a & $\varphiup_\textrm{ba}$ & $\varphiup_\textrm{da}$& $\varphiup_\textrm{ca}$ & $\varphiup_\textrm{fa}$ & main axis \\
         \hline \hline
         A$_\textrm{1g}$ & 0.7 & - & 0.4 & 0.2 & 100° & - & 17° & 228° & 80° \\
         VII & 0.8 & - & 0.3 & 0.1 & 110° & - & 280° & 85° & 166° \\
         V   & 0.5 & - & 0.5 & 0.3 & 0° & - & 280° & 252° & 172° \\
         IV  & 0.4 & 0.5 & - & - & 180° & 268° & - & - & 165° \\
         III & 0.5 & 0.5 & - & - & 179° & 268° & - & - & 161° \\ 
         II  & 0.8 & 0.4 & - & - & 174° & 269°  & - & - & 67° \\ 
         I   & 0.9 & - & 0.1 & 0.1 & 88° & - & 320° & 140° & 36° \\
     \end{tabular}
 \end{subtable}
 \caption{
    Fitting parameters for the polar intensity dependence of RS peaks measured under $P$=7.4 GPa for $\lambdaup$=633 nm (left) and $\lambdaup$=561 nm (right) excitation. The parameter d corresponds to the symmetric case, while c and f to the non-symmetric.
    }
\label{Tbs: SII}    
\end{table}


\bibliographystyle{apsrev4-2}
\bibliography{biblio}